\documentclass[10pt]{article}
\usepackage{amssymb}

\begin{document}

\begin{titlepage}
\begin{flushright}
SPIN-09/46, ITP-UU-09/56
\end{flushright}

\vspace{0.5cm}

\begin{center}
{\Large\bf The Newtonian Limit of Hermitian Gravity}
\end{center}

\vspace{0.3cm}

\begin{center}
{\large Jeroen G.~Burgers $^{\star *}$, Christiaan L.~M.~Mantz $^{\bullet *}$ 
and Tomislav Prokopec $^{\star *}$}
\end{center}
\begin{center}
$^{\star}$ \it{Institute for Theoretical Physics \& Spinoza Institute,
Utrecht University,\\
Leuvenlaan 4, Postbus 80.195, 3508 TD Utrecht, The Netherlands}
\end{center}

\begin{center}
$^{\bullet}$ \it{Free University, Amsterdam\\
The Netherlands}
\end{center}

\vspace{0.2cm}

\vspace{0.3cm}

\begin{center}
ABSTRACT

\end{center}
\hspace{0.3cm} 

 We construct the gauge invariant potentials of Hermitian 
Gravity~\cite{Mantz:2008hm} and derive the  
linearized equations of motion they obey.
A comparison reveals a striking similarity
to the Bardeen potentials of general relativity. We then consider the
response to a point particle source, and discuss in what sense 
the solutions of Hermitian Gravity reduce to the Newtonian potentials. 
In a rather intriguing way, the 
Hermitian Gravity solutions exhibit a generalized
reciprocity symmetry originally proposed by Born in the 1930s. 
Finally, we consider the trajectories of massive and massless particles under
the influence of a potential. The theory correctly reproduces the Newtonian
limit in three dimensions and the nonrelativistic acceleration equation. 
However, it differs from the light deflection calculated in linearized general
relativity by 25\%. While the specific complexification of general relativity 
by extension to Hermitian spaces performed here does not agree with experiment, it does
possess useful properties for quantization and is well-behaved around
singularities as described in~\cite{Mantz:2008hm}. Another form of complex
general relativity may very well agree with experimental data.

\vspace{0.3cm}

\vspace{0.1cm}

\leftline{{\tt pacs:}
%
%
04.50.Kd Modified theories of gravity;
04.20.-q Einstein equation, general relativity
}

\begin{flushleft}
$^{*}$ E-mail: Jeroen.G.Burgers@gmail.com, 
       clmmantz@gmail.com,
       T.Prokopec@uu.nl
\end{flushleft}

\end{titlepage}

\section{Introduction}

 In this paper we consider the dynamics of scalar potentials 
in linearized Hermitian Gravity (HG) recently proposed in 
Ref.~\cite{Mantz:2008hm}. HG enlarges the symmetry
of general relativity (GR) by formulating gravity as 
a geometric theory on phase space $\{x^\mu, p^\mu\}$, 
where $\mu=0,1,..,D-1$, and $D$ is the number of space-time dimensions.
Hermitian Gravity is a generalization of general relativity formulated 
on an eight dimensional phase space $\{x^\mu,p^\nu\}$, where 
$x^\mu$ and $p^\nu$ are {\it a priori} mutually independent coordinates.
Arguably, the simplest realization of such an idea
are curved Hermitian complex spaces in D complex dimensions, 
where $z^\mu = (x^\mu + \imath y^\mu)/\sqrt{2}$
and $z^{\bar \mu} = (x^\mu - \imath y^\mu)/\sqrt{2}$ 
are the holomorphic and antiholomorphic 
coordinates~\footnote{The momentum coordinates $p^{\mu}$ are related to $y^{\mu}$ through the relation $y^\mu = (G_N/c^3)p^\mu$.}   on the (Hermitian) manifold ${\cal M}$ 
with the distance function
\begin{equation}
 ds^2 = C_{\mu\bar \nu}dz^\mu dz^{\bar \nu}
        + C_{\bar \mu\nu}dz^{\bar \mu} dz^\nu
\,.
\label{ds2}
\end{equation}
This line element (and the corresponding metric tensor $C$) 
is invariant under the action of 
the almost complex structure operator $J$:
\begin{equation}
J[ds^2]=ds^2
\,.
\label{Jds}
\end{equation}
The operator $J$ is defined by 
\begin{equation}
J\left[dz^{\mu}\right]=-\imath dz^{\mu}
\,,\qquad
 J\left[d{z}^{\bar\mu}\right]=\imath d{z}^{\bar\mu}
\,,
\label{J}
\end{equation}
where 
\begin{equation}
dz^{\mu}=\frac{1}{\sqrt{2}}\left(dx^{\mu} + \imath dy^{\mu}\right)
\,, \qquad
 d{z}^{\bar\mu}=\frac{1}{\sqrt{2}}\left(dx^{\mu} - \imath dy^{\mu}\right),
\qquad dy^\mu = \frac{G_N}{c^3}dp^\mu
\,.
\label{dz-dzbar}
\end{equation}
Hermitian spaces are complex curved spaces ${\cal M}$ 
endowed with a Hermitian metric $C$ which obeys~(\ref{J}).
As a consequence of the $J$-symmetry~\footnote{In the Hermitian Gravity theory
considered in this paper, the $J$-symmetry and reciprocity symmetry have
identical meaning, and hence will be used interchangeably.}
of the line element~(\ref{Jds}),
the holomorphic and antiholomorphic metric components of the metric tensor 
vanish, $C_{\mu\nu}=C_{\bar{\mu}\bar{\nu}}=0$. One can achieve that
{\it e.g.} by adding a suitable constraint 
action~\footnote{See Eq.~(\ref{Action}) for a particular realization
of the constraint action.}. The reciprocity symmetry is then imposed at the 
level of the equations of motion (on-shell). Equivalently, one can 
solve the constraint equations $C_{\mu\nu}=0$, insert the solution 
into the action, to obtain an effective action, which has no dependence 
on $C_{\mu\nu}$ (but it contains dependence on a Lagrange multiplier tensor).

 The symmetry of HG is indeed much larger than
the diffeomorphism invariance of GR, 
in that the Hermitian line element~(\ref{ds2}) is invariant 
under arbitrary (holomorphic) complex coordinate transformations, 
$z^\mu \rightarrow \tilde z^\mu(z^\rho)
       = (\partial \tilde z^\mu(z^\rho)/\partial z^\alpha) dz^\alpha$,
and it reduces to diffeomorphism invariance in the low energy limit,
$p^\mu\rightarrow 0$. 

 Since GR is an extremely successful and well tested theory, 
the natural question that arises is in what sense GR needs to be improved,
and why should Hermitian Gravity be the desired fix.
 In order to begin answering this question, we recall that the main motivation 
for HG dates back to an old idea of Max Born~\cite{Born1938Reciprocity}. 
Albeit Born's presentation of his reciprocity symmetry is in places unclear,
from his papers one can conclude the following. 
Born was inspired by the symmetry of Hamilton's equations~\footnote{In
modern mathematical language, the Hamiltonian dynamics leaves 
the symplectic structure -- defined as the two form $dx^i\wedge dp^j$ --
invariant.}
\begin{equation}
 dx^i = \frac{\partial H}{\partial p_i} d\tau
\,,\qquad 
 dp^i = -\frac{\partial H}{\partial x_i}d\tau
\,,
\nonumber
\end{equation}
where $\tau$ denotes an affine time. These equations are invariant under
the following simultaneous transformation of the cotangent phase space
at a point, 
\begin{equation}
 \frac{\partial}{\partial x_i}\rightarrow \frac{\partial}{\partial p_i}
\,,\qquad 
 \frac{\partial}{\partial p_i}\rightarrow -\frac{\partial}{\partial x_i}
\label{reciprocity symmetry:cotangent}
\end{equation}
and of the tangent phase space, 
\begin{equation}
 d x^i\rightarrow dp^i
\,,\qquad 
 d p^i\rightarrow - dx^i
\,.
\label{reciprocity symmetry:tangent}
\end{equation}
Notice further that the canonical quantization relation,
\begin{equation}
\left[\hat x^i,\hat p_j\right]\equiv \hat x^i\hat p_j-\hat p_j\hat x^i
        =\imath \hbar \delta^i_j
\label{canonical quantization}
\end{equation}
also obeys reciprocity symmetry in the following sense. 
Let us now rewrite the operators 
$\hat x^i$ and $\hat p_j$ as a sum of their classical values
$x^i \equiv \langle \Omega |\hat x^i|\Omega \rangle$,  
$p_j \equiv \langle \Omega |\hat p_j|\Omega \rangle$,
where $|\Omega\rangle$ denotes quantum state, 
and their fluctuating quantum parts $\{\delta \hat x^i,\delta \hat p_j\}$ as, 
$\hat x^i=x^i+\delta \hat x^i$ and $\hat p_j=p_j+\delta \hat p_j$.
We can now rewrite~(\ref{canonical quantization}) in terms of 
the fluctuating parts only as, 
\begin{equation}
\left[\delta\hat x^i,\delta\hat p_j\right] = \imath \hbar \delta^i_j
\,.
\label{canonical quantization:2}
\end{equation}
From this we immediately see that the fluctuating parts
 $\{\delta \hat x^i,\delta \hat p_j\}$ obey reciprocity symmetry,

\begin{equation}
 \delta\hat x^i\rightarrow \delta\hat p^i
\,,\qquad 
 \delta\hat p^i\rightarrow - \delta\hat x^i
\,,
\label{reciprocity symmetry:quantum}
\end{equation}
which in form (but only in form!)
resembles~(\ref{reciprocity symmetry:tangent}).
The meaning of~(\ref{reciprocity symmetry:quantum}) is however quite different
from that of the classical symmetry~(\ref{reciprocity symmetry:tangent}).
Indeed, Eq.~(\ref{reciprocity symmetry:quantum}) tells us that
canonical quantization is such that the resulting quantum fluctuations
obey reciprocity symmetry. 
This realization is particularly exciting, since it gives hope that 
the quantum theory of gravity might possess reciprocity symmetry.
If true, this would elevate reciprocity symmetry 
to a fundamental symmetry of Nature, and it would help us in finding out 
how to correctly quantize gravity.
Born has realised that the reciprocity 
transformation~(\ref{reciprocity symmetry:tangent}) 
maps low energy (sub-Planckian) physics
to high energy (super-Planckian) physics, and {\it vice versa}.
Since it was not clear how to make the action of matter, 
force, or gravitational fields invariant under such a symmetry,
Born's idea has not attracted much attention.
In Ref.~\cite{Mantz:2008hm}, it was realized that Hermitian spaces 
$({\cal M},C)$, manifolds endowed with complex metrics, 
with the identification~(\ref{dz-dzbar}), 
represent a natural generalization of Born's reciprocity symmetry,
whereby the 3-dimensional 1-forms and vectors 
in~(\ref{reciprocity symmetry:tangent}) 
and~(\ref{reciprocity symmetry:cotangent}), respectively,
are replaced by the 4-vectors of Eqs.~(\ref{J}--\ref{dz-dzbar}),
where by a suitable rescaling the dimensions of $x^\mu$ and $p^\mu$ are made
equal. Just as in Eq.~(\ref{reciprocity symmetry:tangent}),
the Hermitean space reciprocity symmetry is  
realized on one-forms 
\begin{equation}
J[dx^\mu] = \frac{G_N}{c^3}dp^\mu
\,,\qquad 
J[dp^\mu] = -\frac{c^3}{G_N}dx^\mu
\,.
\label{J:dX-dp}
\end{equation}
Analogous relations hold for the vector fields,  
\begin{eqnarray}
J\left[\frac{\partial}{\partial z^{\mu}}\right]
  =\imath \frac{\partial}{\partial z^{\mu}}
\,,\qquad
 J\left[\frac{\partial}{\partial z^{\bar\mu}}\right]
  = - \imath \frac{\partial}{\partial z^{\bar\mu}}\nonumber
\,,
\label{J:vector}
\end{eqnarray}
which imply 
$J[\partial/\partial x^{\mu}] = \partial/\partial y^{\mu}$ and
$J[\partial/\partial y^{\mu}] = -\partial/\partial x^{\mu}$.

 It it convenient to introduce an eight-dimensional 
notation~\cite{Mantz:2008hm}, which we indicate by bold quantities
and Latin indices which run from $0$ to $2D-1$. 
In eight-dimensional notation the
line element~(\ref{ds2}) becomes simply
\begin{equation}
\mbox{\boldmath $ds^2=dz^m C_{mn} z^n\equiv dz\cdot C\cdot d z$}, 
\qquad \mbox{\boldmath $z^m$}=\left(z^{\mu},{z}^{\bar\mu}\right)^{T}
\,.\nonumber
\end{equation}
The eight-dimensional metric
\begin{eqnarray}
{\mathbf C} = \left(\begin{array}{cc}
                       0 & C_{{\mu}\bar\nu} \cr
                      C_{\bar{\mu}\nu} & 0 \cr
                  \end{array}\right),
\,
\label{C:8dim}\nonumber
\end{eqnarray}
is symmetric under transposition,  $\mbox{\boldmath $C_{mn}=C_{nm}$}$, 
while the four-dimensional sectors are Hermitian, $C^\dag=C$, which implies
the off-diagonal sectors are related through complex conjugation: 
$C_{\bar{\mu}\nu}=C_{\mu\bar\nu}^*$ and $C_{\bar{\mu}\nu}=C_{\nu\bar\mu}$. 
By variation of the Hermitian line element, one can derive the Hermitian 
geodesic equations from which the Christoffel connection follows
\begin{equation}
\mbox{\boldmath$\Gamma^r_{mn}$}=\mbox{\boldmath$\frac{1}{2}C^{rs}\left(\partial_m C_{sn} + \partial_n C_{ms} - \partial_s C_{mn}\right)$},
\label{Christoffel}
\end{equation}
while the Riemann tensor is given by
\begin{equation}
\mbox{\boldmath$R^s_{mln}$}=\mbox{\boldmath$\partial_l\Gamma^s_{nm}-\partial_n\Gamma^s_{lm}+\Gamma^s_{la}\Gamma^a_{nm}-\Gamma^s_{na}\Gamma^a_{lm}$}.\nonumber
\end{equation}
The Christoffel symbols and Riemann tensor in eight dimensions satisfy the same symmetries as in general relativity, as can readily be seen by interchanging indices. It is straightforward to calculate the Einstein tensor
\begin{eqnarray}
\mbox{\boldmath$G_{mn}=R_{mn}-\frac{1}{2}C_{mn}R$}.\nonumber
\end{eqnarray}
The Ricci tensor $\mbox{\boldmath$R^r_{\;mrn}\equiv R_{mn}$}$ 
and Ricci scalar 
$\mbox{\boldmath$R\equiv R^m_{\;m}$}=\mbox{\boldmath$\zeta^{mn}R_{nm}$}$ 
are procured through the proper contractions of the Riemann tensor. 

 The paper is organized as follows. 
In section~\ref{Linearized Hermitian Gravity} we calculate the linearized 
Einstein tensor. In section~\ref{Bardeen Potentials} 
and in Appendix A we show how to construct the gauge invariant Bardeen
potentials of Hermitian Gravity and derive the corresponding 
linearized vacuum equations. 
In section~\ref{Newtonian Limit} we solve 
for the potentials of a point static mass both in two and three
(complex) spatial dimensions. Requiring reciprocity 
symmetry brings us naturally to phase space potentials. The Newtonian 
limit of GR is recovered by integrating phase space potentials 
over the momenta. Finally, in section 5 we consider the motion of a
spinless massive test particle and of a massless particle (photon).

\section{Linearized Hermitian Gravity}
\label{Linearized Hermitian Gravity}

Just as in GR, a first step towards the Newtonian limit is 
to linearize the theory. For simplicity we shall 
linearize around the flat Minkowski metric \mbox{\boldmath $\zeta_{mn}$} 
such that the metric \mbox{\boldmath $C_{mn}$} decomposes as
\begin{eqnarray}
\mbox{\boldmath $C_{mn}=\zeta_{mn}+H_{mn}\left(z^r\right)$},\nonumber
\end{eqnarray}
where $\mbox{\boldmath{$\zeta^{mn}$}}$ is an $8\times 8$
symmetric metric whose $4\times 4$ blocks  are, 
\begin{equation}\
\mbox{\boldmath{$\zeta^{mn}$}}=
      \left(\begin{array}{cc}\zeta^{\mu\nu} & \zeta^{\mu\bar\nu} \cr
                   \zeta^{\bar\mu\nu} & \zeta^{\bar\mu\bar\nu} \cr
      \end{array}
\right)
\,,
\end{equation}
where $\zeta^{\bar\mu\nu}={\rm diag}(-1,1,1,1)=\zeta^{\mu\bar\nu}$
and $\zeta^{\mu\nu}={\rm diag}(0,0,0,0)=\zeta^{\bar\mu\bar\nu}$,
and \mbox{\boldmath $H_{mn}$} is 
a small space-time-momentum-energy dependent contribution, 
 $\|\mbox{\boldmath $H_{mn}$}\|\ll 1$.

Upon keeping only terms up to linear order in \mbox{\boldmath $H_{mn}$},
the Christoffel symbols~(\ref{Christoffel}) simplify to, 
\begin{eqnarray}\label{Christoffel2}
\Gamma^\rho_{\mu\nu}&=&\frac{1}{2}C^{\bar{\lambda}\rho}\left(\partial_{\mu} C_{\nu\bar{\lambda}} + \partial_{\nu} C_{\mu\bar{\lambda}} \right) = \frac{1}{2}\zeta^{\bar{\lambda}\rho}\left(\partial_{\mu} H_{\nu\bar{\lambda}} + \partial_{\nu} H_{\mu\bar{\lambda}} \right),\nonumber\\
\Gamma^\rho_{\bar{\mu}\nu}&=&\frac{1}{2}C^{\bar{\lambda}\rho}\left(\partial_{\bar{\mu}} C_{\nu\bar{\lambda}} - \partial_{\bar{\lambda}} C_{\nu\bar{\mu}} \right) = \frac{1}{2}\zeta^{\bar{\lambda}\rho}\left(\partial_{\bar{\mu}} H_{\nu\bar{\lambda}} - \partial_{\bar{\lambda}} H_{\nu\bar{\mu}} \right),\nonumber\\
\Gamma^\rho_{\mu\bar{\nu}}&=&\frac{1}{2}C^{\bar{\lambda}\rho}\left(\partial_{\bar{\nu}} C_{\mu\bar{\lambda}} - \partial_{\bar{\lambda}} C_{\mu\bar{\nu}} \right) = \frac{1}{2}\zeta^{\bar{\lambda}\rho}\left(\partial_{\bar{\nu}} H_{\mu\bar{\lambda}} - \partial_{\bar{\lambda}} H_{\mu\bar{\nu}} \right),\nonumber\\
\Gamma^{\bar{\rho}}_{\bar{\mu}\bar{\nu}}&=&\frac{1}{2}C^{\bar{\rho}\lambda}\left(\partial_{\bar{\mu}} C_{\lambda\bar{\nu}} + \partial_{\bar{\nu}} C_{\lambda\bar{\mu}} \right) = \frac{1}{2}\zeta^{\bar{\rho}\lambda}\left(\partial_{\bar{\mu}} H_{\lambda\bar{\nu}} + \partial_{\bar{\nu}} H_{\lambda\bar{\mu}} \right),\nonumber\\
\Gamma^{\bar{\rho}}_{\mu\bar{\nu}}&=& \frac{1}{2}C^{\bar{\rho}\lambda}\left(\partial_{\mu} C_{\lambda\bar{\nu}} - \partial_{\lambda} C_{\mu\bar{\nu}} \right) = \frac{1}{2}\zeta^{\bar{\rho}\lambda}\left(\partial_{\mu} H_{\lambda\bar{\nu}} - \partial_{\lambda} H_{\mu\bar{\nu}} \right),\nonumber\\
\Gamma^{\bar{\rho}}_{\bar{\mu}\nu}&=& \frac{1}{2}C^{\bar{\rho}\lambda}\left(\partial_{\nu} C_{\lambda\bar{\mu}} - \partial_{\lambda} C_{\nu\bar{\mu}} \right) = \frac{1}{2}\zeta^{\bar{\rho}\lambda}\left(\partial_{\nu} H_{\lambda\bar{\mu}} - \partial_{\lambda} H_{\nu\bar{\mu}} \right),\nonumber\\
\Gamma^{\rho}_{\bar{\mu}\bar{\nu}}&=&0,\nonumber\\
\Gamma^{\bar{\rho}}_{\mu\nu}&=&0.
\end{eqnarray}
Introducing the Christoffel symbols into the Riemann tensor it is straightforward to calculate the Einstein tensor. 
In terms of holomorphic and anti-holomorphic components, the various components
of the Einstein tensor are given by
\begin{eqnarray}
G_{\mu\nu}&=&\frac{1}{2}\left(\partial_{\mu}\partial^{\bar{\lambda}}\tilde{H}_{\bar{\lambda}\nu}+\partial_{\nu}\partial^{\bar{\lambda}}\tilde{H}_{\bar{\lambda}\mu}\right),\nonumber\\
G_{\bar{\mu}\bar{\nu}}&=&\frac{1}{2}\left(\partial_{\bar{\mu}}\partial^{\lambda}\tilde{H}_{\lambda\bar{\nu}}+\partial_{\bar{\nu}}\partial^{\lambda}\tilde{H}_{\lambda\bar{\mu}}\right),\nonumber\\
G_{\bar{\mu}\nu}&=&\frac{1}{2}\partial_{\bar{\mu}}\partial^{\bar{\lambda}}\tilde{H}_{\bar{\lambda}\nu}+\frac{1}{2}\partial_{\nu}\partial^{\rho}\tilde{H}_{\rho\bar{\mu}}-\frac{1}{2}\Box \tilde{H}_{\nu\bar{\mu}}-\zeta_{\bar{\mu}\nu}\partial^{\rho}\partial^{\bar{\sigma}}\tilde{H}_{\bar{\sigma}\rho},\nonumber\\
G_{\mu\bar{\nu}}&=&\frac{1}{2}\partial_{\mu}\partial^{\lambda}\tilde{H}_{\lambda\bar{\nu}}+\frac{1}{2}\partial_{\bar{\nu}}\partial^{\bar{\rho}}\tilde{H}_{\bar{\rho}\mu}-\frac{1}{2}\Box \tilde{H}_{\bar{\nu}\mu}-\zeta_{\mu\bar{\nu}}\partial^{\rho}\partial^{\bar{\sigma}}\tilde{H}_{\bar{\sigma}\rho},
\label{EinTensor}
\end{eqnarray}
where the trace-reversed metric 
perturbation~\footnote{In the case of Hermitian Gravity, 
the trace-reversed perturbation satisfies 
$\mbox{\boldmath$\tilde{H}=-3H$}$.}, $\mbox{\boldmath$\tilde{H}_{mn}=H_{mn}-\frac{1}{2}\zeta_{mn}H$}$, has been employed to simplify the equations and the d' Alembertian is $\Box = - 2\partial_0\partial_{\bar 0}+\nabla^2$ (the Laplacian is defined as
$\nabla^2 = \sum_{i=1}^3 2\partial_i\partial_{\bar i}$).  
The Hermitian-Einstein equation
\begin{equation}
G_{\mu\bar\nu} =  \kappa T_{\mu\bar\nu},
\label{Hermitean-Einstein}
\end{equation}
describes the dynamics of the system.
The constant $\kappa$, which in GR equals $8\pi G_N$ in units where $c=1$,
is determined in section~\ref{Three-Dimensional Case}  below
by requiring that HG reproduces the correct Newtonian limit. 
Equation~(\ref{Hermitean-Einstein}) can be obtained by varying \
the action 
\begin{eqnarray}
S &=& S_{HG} + S_c + S_m
\nonumber\\
S_{HG}&=&\frac{1}{2\kappa}\int \mbox{\boldmath{$d^4zd^4\bar z$}}
      \sqrt{\mbox{\boldmath{$C$}}} \mbox{\boldmath{$R$}} 
\,,\qquad 
S_c = \frac{1}{2\kappa}\int \mbox{\boldmath{$d^4zd^4\bar z$}}
      \sqrt{\mbox{\boldmath{$C$}}} 
             (\lambda_{\mu\nu} C^{\mu\nu}
              +\lambda_{\bar\mu\bar\nu} C^{\bar\mu\bar\nu})
\,,
\label{Action}
\end{eqnarray}
where $\mbox{\boldmath $C$}={\rm det}[\mbox{\boldmath $C_{mn}$}]$,
$S_c$ is a constraint action and $S_m$ is the matter field action.
Notice that varying the constraint action yields $C_{\mu\nu}=0$,
which imposes reciprocity symmetry at the level of the equations of motion 
(on shell). 
This way $\mbox{\boldmath{$R$}}=\mbox{\boldmath{$R$}}[C_{\mu\bar\nu}]$ 
is not a function of $C_{\mu\nu}$. 
In this case, $G_{\mu\nu}$ and  $G_{\bar\mu\bar\nu}$ 
in general do not vanish. 
The Einstein tensor does however obey the Bianchi identity,
$\mbox{\boldmath$\nabla^mG_{mn}=0$}$, and so does the constraint tensor,
$\mbox{\boldmath$\nabla^m\lambda_{mn}=0$}$, which implies
that the constraint tensor must be covariantly conserved,
$\nabla^\mu\lambda_{\mu\nu}=0=\nabla^{\bar\mu}\lambda_{\bar\mu\bar\nu}$.
The action~(\ref{Action}) is constructed in such a way that the number 
of undetermined field components equals the number of equations, 
which is a necessary condition for mathematical consistency of the theory. 
Another possibility is Holomorphic Gravity, in which 
a contraint action analogous to~(\ref{Action})
imposes on-shell constraints $C_{\mu\bar\nu}=C_{\bar\mu\nu}$. This theory
has been studied in some detail in Ref.~\cite{Mantz:2008fj}.

\section{Bardeen Potentials}
\label{Bardeen Potentials}

When looking at linearized theory in general relativity, 
it is useful to decompose the perturbations in terms of the scalar (S), 
vector (V), and tensor (T) perturbations such 
that~\cite{Mukhanov:1990me}
\begin{eqnarray}
h_{\mu\nu}=h_{\mu\nu}^S+h_{\mu\nu}^V+h_{\mu\nu}^T
\,.\nonumber
\end{eqnarray}
In studying what kind of potential is generated by a point mass, 
it is only necessary to consider scalar perturbations which correspond
 to fluctuations of mass density with respect to the background field. 
Vector and tensor perturbation can be produced {\it e.g.} by  
amplified vacuum fluctuations of vector and gravitational fields in inflation.
In macroscopic systems, vector perturbation are sourced  
by a time dependent dipole of some localized mass distribution, 
while tensor perturbations 
(gravitational waves) are sourced by a time dependent quadrupole 
of a mass distribution. Typical astronomical sources of vector and tensor
perturbations are rotating binary stars. For the scalar perturbations, 
the perturbation metric can be expressed in terms of the scalar fields
$E, B, \phi$ and $\psi$
\begin{eqnarray}
h_{\mu\nu}^S=\left(\begin{array}{cccc}
-2\phi&\partial_1 B&\partial_2 B&\partial_3 B\\
\partial_1 B&2\left(-\psi+\partial_1\partial_1 E\right)&2\partial_1\partial_2 E&2\partial_1\partial_3 E\\
\partial_2 B&2\partial_2\partial_1 E&2\left(-\psi+\partial_2\partial_2 E\right)&2\partial_2\partial_3 E\\
\partial_3 B&2\partial_3\partial_1 E&2\partial_3\partial_2 E&2\left(-\psi+\partial_3\partial_3 E\right)\end{array}\right).\nonumber
\end{eqnarray}
The fields that make up the metric perturbation are not invariant under gauge
transformations, {\it i.e.} coordinate transformations generated by
$x^{\mu}\rightarrow x^{\mu} + \xi^{\mu}(x^{\nu})$, where $\xi^{\mu}(x^{\nu})$ 
is an infinitessimal vector field~\footnote{For the case of scalar 
perturbations, $\xi^{\mu}(x^{\nu})=\left(\xi^0, \partial_i \xi\right)^T$.}. 
One can show that it is possible to construct so-called gauge invariant 
potentials. The advantage of a gauge invariant formulation is that the
remaining potentials represent physical quantities invariant under 
diffeomorphisms. In general relativity for a flat background metric, 
the gauge invariant potentials, also known as Bardeen potentials, are given by
\begin{equation}\label{Bardeen:pot}
\Phi_{GR} = \phi + \partial_0\left(B-\partial_0 E\right),\qquad
\Psi_{GR} = \psi
\,.
\end{equation}
One can introduce an analogous construction in Hermitian Gravity which satisfies the symmetries of the theory. An analysis of these symmetries indicates the $E, \phi,$ and $\psi$ fields must be real, while the field $B$ is not imposed with any such constraint and thus has its own complex conjugate $\bar{B}$. 
Under these conditions, 
the Hermitian Gravity analog of the Bardeen decomposition for the scalar 
metric perturbation can be constructed:
\begin{eqnarray}\label{bardpotmetric}
H_{\bar{\mu}\nu}^S=\left(\begin{array}{cccc}
-2\phi&\partial_1 B&\partial_2 B&\partial_3 B\\
\partial_{\bar{1}} \bar{B}&2\left(-\psi+\partial_{\bar{1}}\partial_1 E\right)&2\partial_{\bar{1}}\partial_2 E&2\partial_{\bar{1}}\partial_3 E\\
\partial_{\bar{2}} \bar{B}&2\partial_{\bar{2}}\partial_1 E&2\left(-\psi+\partial_{\bar{2}}\partial_2 E\right)&2\partial_{\bar{2}}\partial_3 E\\
\partial_{\bar{3}} \bar{B}&2\partial_{\bar{3}}\partial_1 E&2\partial_{\bar{3}}\partial_2 E&2\left(-\psi+\partial_{\bar{3}}\partial_3 E\right)\end{array}\right).
\end{eqnarray}
Under an infinitesimal transformation
$\mbox{\boldmath$z^m\rightarrow z^m + \xi^m(z^n)$}$, the metric perturbation 
transforms as follows 
\begin{equation}
\mbox{\boldmath$H_{mn}\left(z^m\right)
\rightarrow H'_{mn}=H_{mn}-\partial_m\xi_n-\partial_n\xi_m$}
\,,\qquad \mbox{\boldmath$\xi_n = \zeta_{mn}\xi^m$}
\,.
\label{gauge transformation}
\end{equation}
The coordinate transformation consists
of a holomorphic and an antiholomorphic vector field \
$\mbox{\boldmath$\xi^m$}
 =\left(\xi^{\mu}(z^{\nu}),\bar{\xi}^{\mu}(z^{\bar{\mu}})\right)^T$. 
This is a condition which must be imposed in order to keep the metric
Hermitian ({\it i.e.} to maintain $C_{\mu\nu}=0$ and $C_{\bar{\mu}\bar{\nu}}=0$)
under coordinate transformations.
Equation~(\ref{gauge transformation}) 
implies the following gauge transformations for the scalar
fields~\footnote{For scalar perturbations, one can use $\xi^i=\partial_i \xi$.}
\begin{eqnarray}\label{gaugetr}
2\phi&\rightarrow& 2\phi - \partial_0 \xi^0 - \partial_{\bar{0}}\xi^{\bar{0}},\nonumber\\
-\partial_{\bar{\i}}\bar{B}&\rightarrow& -\partial_{\bar{\i}}\bar{B} + \partial_0 \partial_i \xi - \partial_{\bar{\i}} \xi^{\bar{0}},\nonumber\\
-\partial_i B &\rightarrow& -\partial_i B + \partial_{\bar{0}}\partial_{\bar{\i}}\bar{\xi} - \partial_i \xi^0,\nonumber\\
\psi&\rightarrow& \psi,\nonumber\\
-2 \partial_i \partial_{\bar{\j}} E &\rightarrow& -2 \partial_i \partial_{\bar{\j}} E + \partial_i \partial_j \xi + \partial_{\bar{\i}}\partial_{\bar{\j}} \bar{\xi}.
\end{eqnarray}
Computationally, it is much simpler to work with the trace-reversed 
metric which was introduced in the Einstein
equations~(\ref{EinTensor}--\ref{Hermitean-Einstein}). 
The new trace-reversed fields $\tilde{\phi}$ and $\tilde{\psi}$ transform
as%
\begin{eqnarray}
2\tilde{\phi}&=&6\psi - \nabla^2 E\longrightarrow 2\tilde{\phi} + \left(\partial_j\right)^2\xi + \left(\partial_{\bar{i}}\right)^2\bar{\xi},\nonumber\\
2\tilde{\psi}&=&-4\psi +2\phi +\nabla^2 E\longrightarrow 2\tilde{\psi} -\partial_0 \xi^0 - \partial_{\bar{0}}\xi^{\bar{0}}-\left(\partial_k\right)^2\xi - \left(\partial_{\bar{k}}\right)^2\bar{\xi},
\nonumber
\end{eqnarray}
while the two other fields, $E$ and $B$, remain unchanged under trace
reversal.  
Now one can determine the dynamics of the fields are governed by the equations (\ref{EinTensor}) resulting through the choice of energy-momentum tensor as described in (\ref{Hermitean-Einstein}). In preparation for the following section on the Newtonian limit, we will use~\footnote{This corresponds to the energy-momentum tensor of a perfect fluid in the rest frame.} $T_{\bar{\mu} \nu}={\rm diag}\left(-\rho,0,0,0\right)$, so that the equations in (\ref{EinTensor}) simplify to
\begin{eqnarray}\label{eomrev1}\label{boxfields}
\Box 2\tilde{\phi}&=&\frac{1}{2}\left(\partial_0 \nabla^2 B + \partial_{\bar{0}} \nabla^2 \bar{B}\right) + 2\nabla^2 \tilde{\psi} - \nabla^4 E+2\kappa \rho,\nonumber\\
\Box \partial_i B &=&-\partial_0 \partial_{\bar{0}} \partial_i B- 2 \partial_{\bar{0}} \partial_i \tilde{\psi} + \partial_{\bar{0}} \partial_i \nabla^2 E + 2 \partial_{\bar{0}} \partial_i \tilde{\phi} + \frac{1}{2}\partial_i \nabla^2 B,\nonumber\\
\Box \partial_{\bar{i}} \bar{B} &=&-\partial_0 \partial_{\bar{0}} \partial_{\bar{i}} \bar{B}- 2 \partial_0 \partial_{\bar{i}} \tilde{\psi} + \partial_0 \partial_{\bar{i}} \nabla^2 E + 2 \partial_0 \partial_{\bar{i}} \tilde{\phi} + \frac{1}{2}\partial_{\bar{i}} \nabla^2 \bar{B},\nonumber\\
2\Box \left( \tilde{\psi} \delta_{i\bar{j}} - \partial_i \partial_{\bar{j}} E\right)&=&\partial_{\bar{0}}\partial_i\partial_{\bar{j}}\bar{B} + \partial_0\partial_i\partial_{\bar{j}} B + 4\partial_i \partial_{\bar{j}}\tilde{\psi} - 2 \partial_i \partial_{\bar{j}} \nabla^2 E \nonumber\\
&&- \delta_{i\bar{j}}\left(4 \partial_0 \partial_{\bar{0}} \tilde{\phi} + \partial_0 \nabla^2 B + \partial_{\bar{0}} \nabla^2 \bar{B} + 2 \nabla^2 \tilde{\psi} - \nabla^4 E\right).\nonumber\\
\end{eqnarray}
In Appendix~A we show that, by making the appropriate combinations 
of the fields, one can construct gauge invariant potentials, namely 
\begin{equation}
\Phi=\phi+\partial_0\left(B-\partial_{\bar{0}}E\right)
\,,\qquad 
\Psi = \psi
\,,
\label{Bardeen potentials:HG}
\end{equation}
which are remarkably similar to the Bardeen potentials of 
GR~(\ref{Bardeen:pot}). 
The gauge invariant potentials of HG~(\ref{Bardeen potentials:HG}) 
satisfy the following equations of motion for the rest frame of a perfect fluid
\begin{eqnarray}
\Box \Phi &=& \Box \left( \phi + \partial_0\left(B-\partial_{\bar{0}}E\right)\right)=\frac{2\kappa \rho}{3},\nonumber\\
\Box \Psi &=& \Box \psi =\frac{\kappa \rho}{3}.\nonumber
\end{eqnarray}
In the next section, a massive static point-like source will be considered
such that $\rho$ takes on the form of a delta function with some mass attached to it.\\
As previously stated, it is not yet certain whether or not one is allowed to fix the parts of the Einstein tensor $G_{\mu\nu}$ and $G_{\bar{\mu}\bar{\nu}}$ to zero. Such a choice imposes that the de Donder gauge condition $\mbox{\boldmath$\partial^m H_{mn}$}$ is also null. However, unlike general relativity, this quantity is gauge invariant in Hermitian Gravity. For more in depth information regarding
this constraint and the resultant calculation for the Bardeen potentials, 
see Appendix B.

\section{Newtonian Limit}
\label{Newtonian Limit}

 Following the introduction of the linearized theory of Hermitian Gravity and the use of Bardeen potentials, the gauge invariant scalar potentials in the case of a non-zero energy density are 
\begin{equation}
\Box \Phi = \frac{2\kappa \rho_{HG}}{3}
\,,\qquad \Box \Psi = \frac{\kappa\rho_{HG}}{3}
\,.
\label{eom:Phi}
\end{equation}
 The constant $\kappa$ has yet to be determined and will be done so by solving (\ref{eom:Phi}) for the case of the static point-like mass for the gauge invariant energy density $\rho_{HG}$. 
In general relativity, this energy density 
is given by $\rho_N = M \delta^3(\vec x\,)$. In Hermitian Gravity, the corresponding general relativistic limit can be obtained by considering the point-like source on the six-dimensional phase space spanned by $\{\vec z,\vec {\bar z}\}$
\begin{eqnarray}
\rho_{HG}=M \delta^3(\vec z)\delta^3(\vec {\bar z}),\nonumber
\end{eqnarray}
and subsequently integrating out the momenta contributions 
\begin{eqnarray}
 \rho_N = M\int d^3 y \delta^3(\vec z\,)\delta^3(\vec {\bar z}\,)
\,.\nonumber
\end{eqnarray}
By taking the non-relativistic limit in $d$ complex dimensions~\footnote{This is equivalent to $2d$ real dimensions.}, 
Eq.~(\ref{eom:Phi}) reduces to
\begin{equation}
\nabla^2 \Phi = \frac{2\kappa}{3} M \delta^d(\vec z\,)\delta^d(\vec {\bar z}\,),
\label{eom:Phi:2}
\end{equation}
where $d=D-1$ represents the number of complex, spatial dimensions~\footnote{$D$ is the number of complex space-time dimensions.}.

\subsection{Two-Dimensional Case}
\label{Two-Dimensional Case}

 The first case to be considered is the $d=2$ dimensional case. The
spherical coordinates on $U(2)$ are obtained by the use of the following coordinates:
\begin{eqnarray}
  z^{1} &=& z \cos(\phi)
            \equiv \frac{1}{\sqrt{2}}R {\rm e}^{\imath\alpha}\cos(\phi)
\,,\qquad (z\in C,\; R\in R_+,\; 0\leq \alpha,\phi < 2\pi)
\nonumber\\
  z^{2} &=& z \sin(\phi) {\rm e}^{\imath\theta}
          \equiv \frac{1}{\sqrt{2}}R {\rm e}^{\imath(\alpha+\theta)}\sin(\phi)
\,,\qquad (0\leq \theta < 2\pi)
\label{coordinate trans:2d}
\end{eqnarray}
and analogously for $z^{\bar 1} = (z^1)^*, z^{\bar 2} = (z^2)^*$.
In order to solve Eq.~(\ref{eom:Phi:2}) in these spherical coordinates,
it will be necessary to find the transformed Laplacian $\nabla^2$ in these coordinates.
This can be achieved as follows.
The coordinate transformation from $\mbox{\boldmath$z^m \rightarrow x^m(z^m)$}$
($\mbox{\boldmath$m$}=1,2,3,4$; $\mbox{\boldmath$z^m$} = (z^i,z^{\bar i})^T$, $\mbox{\boldmath$x^m$}=(R,\phi,\alpha,\theta)^T$)
implies
\begin{eqnarray}
 \mbox{\boldmath$dz^m = \frac{\partial z^m}{\partial x^n}dx^n$}
\nonumber
\end{eqnarray}
which in matrix notation reads
\begin{equation}
 d\vec z = D d\vec x
\,,\qquad  d\vec x = D^{-1} d\vec z
\,,
\label{coordinate transf:2d:2}
\end{equation}
where $D$ is the transformation matrix represented by
\begin{eqnarray}
 D =\frac{1}{\sqrt{2}}\left(\begin{array}{cccc}
        {\rm e}^{\imath\alpha}\cos(\phi) & -R{\rm e}^{\imath\alpha}\sin(\phi) &
        \imath R{\rm e}^{\imath\alpha}\cos(\phi) &        0 \cr
        {\rm e}^{\imath(\alpha+\theta)}\sin(\phi) &
                  R{\rm e}^{\imath(\alpha+\theta)}\cos(\phi) &
        \imath R{\rm e}^{\imath(\alpha+\theta)}\sin(\phi) &
                 \imath R{\rm e}^{\imath(\alpha+\theta)}\sin(\phi) \cr
      {\rm e}^{-\imath\alpha}\cos(\phi) & -R{\rm e}^{-\imath\alpha}\sin(\phi) &
        -\imath R{\rm e}^{-\imath\alpha}\cos(\phi) &        0 \cr
        {\rm e}^{-\imath(\alpha+\theta)}\sin(\phi) &
                  R{\rm e}^{-\imath(\alpha+\theta)}\cos(\phi) &
        -\imath R{\rm e}^{-\imath(\alpha+\theta)}\sin(\phi) &
                 -\imath R{\rm e}^{-\imath(\alpha+\theta)}\sin(\phi) \cr
    \end{array}\right),\nonumber
\,.
\label{matrix D}
\end{eqnarray}
and $D^{-1}$ is the inverse of this matrix satisfying $D^{-1} D = I_4$. The Jacobian of the coordinate transformation~(\ref{coordinate transf:2d:2})
is
\begin{eqnarray}
  J = |{\rm det}[D]| = R^3 |\sin(\phi)\cos(\phi)|
  \,.\nonumber
\label{Jacobian}
\end{eqnarray}
From~(\ref{coordinate transf:2d:2}), one can easily see that
\begin{eqnarray}
 \nabla_{\vec z} = (D^{-1})^T \nabla_{\vec x}
\,,\qquad
 \nabla_{\vec x} = D^T \nabla_{\vec z}
\,.\nonumber
\label{coordinate transf:2d:3}
\end{eqnarray}
The transposed inverse matrix reads 
\begin{eqnarray}
 (D^{-1})^T =\frac{1}{\sqrt{2}}\left(\begin{array}{cccc}
        {\rm e}^{-\imath\alpha}\cos(\phi) &
               -\frac{{\rm e}^{-\imath\alpha}\sin(\phi)}{R} &
               \frac{{\rm e}^{-\imath\alpha}}{\imath R\cos(\phi)} &
               -\frac{{\rm e}^{-\imath\alpha}}{\imath R\cos(\phi)}     \cr
        {\rm e}^{-\imath(\alpha+\theta)}\sin(\phi) &
                  \frac{{\rm e}^{-\imath(\alpha+\theta)}\cos(\phi)}{R} &
                                   0  &
               \frac{{\rm e}^{-\imath(\alpha+\theta)}}{\imath R\sin(\phi)} \cr
        {\rm e}^{\imath\alpha}\cos(\phi) &
               -\frac{{\rm e}^{\imath\alpha}\sin(\phi)}{R} &
               -\frac{{\rm e}^{\imath\alpha}}{\imath R\cos(\phi)} &
               \frac{{\rm e}^{\imath\alpha}}{\imath R\cos(\phi)}     \cr
        {\rm e}^{\imath(\alpha+\theta)}\sin(\phi) &
                  \frac{{\rm e}^{\imath(\alpha+\theta)}\cos(\phi)}{R} &
                                   0  &
               -\frac{{\rm e}^{\imath(\alpha+\theta)}}{\imath R\sin(\phi)} \cr
    \end{array}\right)\nonumber
\,.
\label{matrix D:inverseT}
\end{eqnarray}
Using the above matrix, it is possible to determine the Laplace operator
in~(\ref{eom:Phi:2}) in
spherical $U(2)$ coordinates~(\ref{coordinate trans:2d}) through straightforward calculation.
The result is
\begin{eqnarray}
 \nabla^2 &=& \frac{\partial^2}{\partial R^2}
          + \frac{3}{R}\frac{\partial}{\partial R}
          +\frac{1}{R^2}\Bigg[
                   \frac{\partial^2}{\partial \phi^2}
                       + 2\cot(2\phi)\frac{\partial}{\partial \phi}
\nonumber\\
  && +\, \frac{1}{\cos^2(\phi)}\frac{\partial^2}{\partial \alpha^2}
                  + \frac{1}{\sin^2(\phi)\cos^2(\phi)}
                       \frac{\partial^2}{\partial \theta^2}
                  - \frac{2}{\cos^2(\phi)}
                       \frac{\partial}{\partial \alpha}
                       \frac{\partial}{\partial \theta}
                \Bigg]
\,.
\label{Laplace:2d}
\end{eqnarray}
The Laplacian can alternatively be calculated by using
\begin{equation}\label{fastlaplace}
\nabla^2 = \frac{1}{\sqrt{C}}\partial_i \left(\sqrt{C} \, C^{i\bar{j}} \, \partial_{\bar{j}}\right),
\end{equation}
where $C={\rm det}[C_{i\bar{j}}]$ 
is the determinant of the spatial part of the metric and $C^{i\bar{j}}$ 
is the inverse metric satisfying $C^{i\bar{l}}C_{\bar{l}j}=\delta^i_{\;j}$.
 This relation is valid in general relativity (for $g_{ij}$) and, 
using it for the two dimensional case above, 
one can verify that this analogous construction holds true in Hermitian Gravity. 
This method is computationally much quicker and for this reason 
it will be adopted henceforth for the three-dimensional case. \\

 We shall now show how the Laplacian~(\ref{Laplace:2d})
can be broken into two reciprocally invariant parts. 
But before we do that, let us consider the spatial line element $d\ell^2$
in the angular coordinates ({\it cf.} Eq.~(\ref{coordinate trans:2d}))
 \begin{eqnarray}
 z^1 = z \cos(\phi)\,,\qquad  z^2 = z \sin(\phi){\rm e}^{\imath\theta}
\,,\qquad z = \frac{R}{\sqrt{2}}\,{\rm e}^{\imath\alpha}
\,.\nonumber
\end{eqnarray}
After some simple algebra we obtain, 
\begin{eqnarray}
 {d{\vec \ell}}^{\;\,2} = dzd\bar z  
  + z\bar z \left[(d\phi)^2+\sin^2(\phi)(d\theta)^2\right]
  + \imath \left[zd\bar z-\bar z dz\right]\sin^2(\phi)d\theta
\,.
\label{line element:2dim}
\end{eqnarray}
Since the following relations can be shown to be true
\begin{eqnarray}
 J[dR] &=& R(d\alpha+\sin^2(\phi)d\theta)
\,;\qquad 
 J[R(d\alpha+\sin^2(\phi)d\theta)] = -dR,
\nonumber\\
 J[d\phi] &=& -\sin(\phi)\cos(\phi)d\theta
\;\;;\qquad \quad\;\;
 J[\sin(\phi)\cos(\phi)d\theta] = d\phi,\nonumber
\label{J:dcoordinates}
\end{eqnarray}
equation~(\ref{line element:2dim}) can be broken
into two reciprocally invariant pieces as follows:
\begin{eqnarray}
 {d{\vec \ell}}^{\;\,2} 
  = [dz+\imath z\sin^2(\phi)d\theta][d\bar z-\imath \bar z\sin^2(\phi)d\theta]
  + z\bar z \left[(d\phi)^2+\sin^2(\phi)\cos^2(\phi)(d\theta)^2\right]
\,.\nonumber
\label{line element:2dim:b}
\end{eqnarray}
The first term on the right hand side is the radial part while the second
is the angular reciprocally invariant contribution. Notice that, within
the radial part, the angular and radial coordinate elements $dz,d\bar z$, and 
$d\theta$ are mixed. This is to be contrasted with GR, where 
${d{\vec \ell}}^{\;\,2}_{GR} = dR^2 + R^2(d\phi)^2$ is diagonal 
and no such mixing occurs.

 Just like the line element~(\ref{line element:2dim}),
the Laplacian~(\ref{Laplace:2d}) can also be broken into
two reciprocally invariant parts. This can be shown most readily
with Eq.~(\ref{coordinate trans:2d}) by using the quantity $z=\frac{R}{\sqrt{2}}{\rm e}^{\imath \alpha}$, the complex radius, and its complex conjugate.
The derivatives with respect to these coordinates in terms of the Cartesian derivatives are
\begin{eqnarray}
  \frac{\partial}{\partial z} &=& \cos(\phi)\frac{\partial}{\partial z^1}
                                                + \sin(\phi) {\rm e}^{\imath\theta} \frac{\partial}{\partial z^2},
\nonumber\\
\frac{\partial}{\partial \bar z} &=& \cos(\phi)\frac{\partial}{\partial z^{\bar 1}}
                                                     + \sin(\phi) {\rm e}^{-\imath\theta} \frac{\partial}{\partial z^{\bar 2}},
\nonumber\\
 \frac{\partial}{\partial \theta} &=& \imath z\sin(\phi){\rm e}^{\imath\theta}\frac{\partial}{\partial z^2}
                                                       -\imath\bar z \sin(\phi){\rm e}^{-\imath\theta}\frac{\partial}{\partial z^{\bar 2}},
\label{coordinate trans:zz}\\
 \frac{\partial}{\partial \phi} &=& -z\sin(\phi)\frac{\partial}{\partial z^1}
                                                       -\bar z \sin(\phi) \frac{\partial}{\partial z^{\bar 1}}
                                                       + z\cos(\phi){\rm e}^{\imath\theta}\frac{\partial}{\partial z^2}
                                                       +\bar z \cos(\phi) {\rm e}^{-\imath\theta}                                                        \frac{\partial}{\partial z^{\bar 2}}.
 \nonumber
\end{eqnarray}
Recalling that the almost complex structure operator $J$ acts as
\begin{eqnarray}
  J\Big[\frac{\partial}{\partial z^i}\Big] = \imath \frac{\partial}{\partial z^i}
\,,\qquad
 J\Big[\frac{\partial}{\partial z^{\bar i}}\Big]
        =  - \imath \frac{\partial}{\partial z^{\bar i}},
        \qquad (i=1,2),\nonumber
\end{eqnarray}
one can conclude from Eqs.~(\ref{coordinate trans:zz}) the action of $J$
on $\partial/\partial z$ and $\partial/\partial \bar z$ 
\begin{eqnarray}
   J\Big[\frac{\partial}{\partial z}\Big] = \imath \frac{\partial}{\partial z}
\,,\qquad
 J\Big[\frac{\partial}{\partial \bar z}\Big] =  - \imath \frac{\partial}{\partial \bar z}
 \,.\nonumber
\end{eqnarray}
Furthermore, one can check that $\partial/\partial z$
and $\partial/\partial \bar z$ commute
\begin{eqnarray}
   \Big[\frac{\partial}{\partial z},\frac{\partial}{\partial \bar z}\Big] =  0
   \,.\nonumber
\end{eqnarray}
This then implies that the reciprocally invariant radial Laplacian operator is
\begin{eqnarray}
   \nabla^2_{\rm rad}
   &=& 2\frac{\partial}{\partial z}\frac{\partial}{\partial \bar z}
   =  2\cos^2(\phi)\frac{\partial}{\partial z^1}\frac{\partial}{\partial z^{\bar 1}}
       + 2\sin^2(\phi)\frac{\partial}{\partial z^2}\frac{\partial}{\partial z^{\bar 2}}
\nonumber\\
  &&+\,   \sin(2\phi){\rm e}^{-\imath\theta}
             \frac{\partial}{\partial z^1}\frac{\partial}{\partial z^{\bar 2}}
       + \sin(2\phi){\rm e}^{\imath\theta}
            \frac{\partial}{\partial z^2}\frac{\partial}{\partial z^{\bar 1}}
       \,.
\label{nabla:rad}
\end{eqnarray}
Since the entire Laplacian $\nabla^2$ is reciprocally invariant, the angular part of the Laplacian,
$\nabla^2_{\rm ang}=\nabla^2 - 2\partial_z \partial_{\bar z}$, or equivalently
\begin{eqnarray}
   \nabla^2_{\rm ang}
   &=& 2\sin^2(\phi)\frac{\partial}{\partial z^1}\frac{\partial}{\partial z^{\bar 1}}
       + 2\cos^2(\phi)\frac{\partial}{\partial z^2}\frac{\partial}{\partial z^{\bar 2}}
\nonumber\\
  &&-\,   \sin(2\phi){\rm e}^{-\imath\theta}
             \frac{\partial}{\partial z^1}\frac{\partial}{\partial z^{\bar 2}}
       - \sin(2\phi){\rm e}^{\imath\theta}
            \frac{\partial}{\partial z^2}\frac{\partial}{\partial z^{\bar 1}}
\,,
\label{nabla:ang}
\end{eqnarray}
must be as well. When expressed in terms of the coordinates $\{R,\alpha,\phi,\theta\}$ defined
in Eq.~(\ref{coordinate trans:2d}), the relations~(\ref{nabla:rad}--\ref{nabla:ang})
become
\begin{eqnarray}
 \nabla^2_{\rm rad} &=&  2\frac{\partial}{\partial z}\frac{\partial}{\partial \bar z}
   =  \frac{\partial^2}{\partial R^2}
         + \frac{1}{R}\frac{\partial}{\partial R}
         + \frac{1}{R^2}\frac{\partial^2}{\partial\alpha^2}
\label{nabla:rad:2}\\
\nabla^2_{\rm ang} &=& \frac{2}{R}\frac{\partial}{\partial R}
         + \frac{1}{R^2}\bigg[
             \frac{\partial^2}{\partial\phi^2} + 2\cot(2\phi)\frac{\partial}{\partial \phi}
             +\tan^2(\phi)\frac{\partial^2}{\partial\alpha^2}
\label{nabla:ang:2}\\
           &&  \hskip 2cm
           +\,\frac{1}{\sin^2(\phi)\cos^2(\phi)}\frac{\partial^2}{\partial\theta^2}
             -\frac{2}{\cos^2(\phi)}\frac{\partial}{\partial \alpha}\frac{\partial}{\partial\theta}
             \bigg]
\nonumber
\,.
\end{eqnarray}
The simplest way to derive Eqs.~(\ref{nabla:rad:2}--\ref{nabla:ang:2})
is to use the coordinate transformation
\begin{equation}
z=\frac{R}{\sqrt{2}}{\rm e}^{\imath \alpha}
 = \frac{1}{\sqrt{2}}(r_x+\imath r_y)
\label{coordinte trans:xy}
\end{equation}
and
\begin{eqnarray}
\nabla^2_{\rm rad}=2\frac{\partial}{\partial z}
                    \frac{\partial}{\partial \bar z}
              = \frac{\partial^2}{\partial r_x^2}
                  + \frac{\partial^2}{\partial r_y^2}
\,.\nonumber
\label{Laplacian:first}
\end{eqnarray}
Remarkably, from this form of the radial Laplacian
and the coordinates~(\ref{coordinte trans:xy})
it is clear that the radial reciprocally invariant part of the Laplacian
corresponds to a `flat' space Laplacian of the complex plane defined
by the two dimensional hypersurface,
\begin{equation}
  z^1z^{\bar 1}+z^2z^{\bar 2} = z\bar z
\,.
\label{Hyperplane:1}
\end{equation}
However, the complex plane~(\ref{Hyperplane:1}) is {\it not} a flat 
embedding into the space $\{z^1,z^{\bar 1}, z^2,z^{\bar 2}\}$. This can be 
seen from the angular part of the Laplacian~(\ref{nabla:ang:2}) which 
contains both $\partial/\partial_R$ and   $\partial/\partial_\alpha$ 
derivatives.

The reciprocally invariant radial solutions then correspond
to the solutions of the equations
\begin{eqnarray}
&& \bigg(\frac{\partial^2}{\partial R^2}
         + \frac{1}{R}\frac{\partial}{\partial R}
         + \frac{1}{R^2}\frac{\partial^2}{\partial\alpha^2}\bigg)\Phi
 \;=\; \frac{2\kappa}{3} M\delta^2(\vec z\,)\delta(\vec{\bar z}\,),
\label{nabla:rad:3}\\
&&\bigg(\frac{2}{R}\frac{\partial}{\partial R}
         + \frac{1}{R^2}\bigg[
             \frac{\partial^2}{\partial\phi^2} + 2\cot(2\phi)\frac{\partial}{\partial \phi}
             +\tan^2(\phi)\frac{\partial^2}{\partial\alpha^2}
\label{nabla:ang:3}\\
           &&  \hskip 2.5cm
           +\,\frac{1}{\sin^2(\phi)\cos^2(\phi)}\frac{\partial^2}{\partial\theta^2}
             -\frac{2}{\cos^2(\phi)}\frac{\partial}{\partial \alpha}\frac{\partial}{\partial\theta}
             \bigg]\bigg)\Phi=0
\nonumber
\,.
\end{eqnarray}
These equations represent reciprocally invariant solutions which 
propagate on the reciprocally invariant radial hyperplane~(\ref{Hyperplane:1})
only. While the first equation
is derived from purely radial components, 
it should not be forgotten that the angular invariant equation also contains a $\frac{2}{R}\frac{\partial}{\partial R}$ term, indicating that the $z\bar{z}$ hypersurface cannot be isolated through reciprocal invariance. Rather, it is embedded and curved in the other two dimensions. Thus, the reciprocally invariants equations do not purely split into radial and angular parts. Since we desire to study the Newtonian limit, in which the potential $\Phi$ should be only radially dependent (the gravitational force does not depend on angles), it will be necessary to use the full Laplacian. 

 As a result, it is more beneficial to consider a different problem. First, observe that 
in fact the point mass density source in~(\ref{nabla:rad:3}) corresponds to
a point source in phase space. This can be seen by recalling 
 $\vec z=(1/\sqrt{2})(\vec x+\imath \vec y\,)$, $\vec y=(G_N/c^3)\vec p$ and
\begin{eqnarray}
  \delta^2(\vec z\,)\delta^2(\vec{\bar z}\,) = \delta^2(\vec x\,)\delta^2(\vec y\,),
\nonumber
\end{eqnarray}
such that for the point source density
\begin{equation}
\rho_{HG} = M\delta^2(\vec z\,)\delta^2(\vec{\bar z}\,),
\label{rho:point phase space}
\end{equation}
the general relativistic mass density is obtained by 
integrating~(\ref{rho:point phase space}) over the momenta
\begin{eqnarray}
  \frac{G_N^2}{c^6}\int d^2 p \, \delta^2(\vec z\,)\delta^2(\vec{\bar z}\,)
        = M\delta^2(\vec x\,) \equiv \rho_{\rm GR}.\nonumber
\label{point souce:x}
\end{eqnarray}
In other words, the Hermitian Gravity point source~(\ref{rho:point phase space})
represents a point source in phase space $\{\vec x,\vec p\,\}$, and therefore 
the Hermitian Gravity potential $\Phi$ represents a phase space potential (a distribution function)
from which one obtains the limit of general relativity by integrating $\Phi$ over the momenta.
Before illustrating this procedure in the two-dimensional case, 
consider again the Laplace equation~(\ref{eom:Phi:2}) in Cartesian coordinates,
which in $d=2$ spatial dimensions reads
\begin{equation}
      \Bigg(\frac{\partial^2}{\partial \vec x\,^2}
                  + \frac{\partial^2}{\partial \vec y\,^2}\;\Bigg) \Phi 
            = \frac{2\kappa}{3} M \delta^2(\vec x\,)\delta^2(\vec y\,)
\,.
\label{eom:Phi:xy:1}
\end{equation}
Integrating this over $\vec y$, using Gauss's theorem for the $\vec y$ 
integral, and assuming that the surface term 
$\int dS^1_y (d/dr_y) \Phi = [2\pi r_y (d/dr_y)\Phi]_{r_y\rightarrow \infty}$
($r_y=\|\vec y\,\|$)
does not contribute, the integrated potential
\begin{eqnarray}
\phi_x(\vec x\,) \equiv \int d^2 y \Phi,
\,\nonumber
\label{phi_x:integrated}
\end{eqnarray}
satisfies the following Poisson equation
\begin{equation}
  \nabla_{\vec x}^2\, \phi_x(\vec x) = \frac{2\kappa}{3} M \delta^2(\vec x\,)
\,.
\label{Laplace:GR limit:x}
\end{equation}  
This is just the Newtonian limit of GR in the case $\frac{2\kappa}{3} = 4 \pi G_N$, where $G_N$ is Newton's gravitational constant, and has a well known solution
\begin{equation}
 \phi_x\rightarrow \phi_N = \phi_0 + G_NM\ln\bigg(\frac{r^2}{\mu^2}\bigg)
\,.
\label{potential:Newton:x}
\end{equation}
Here, $\phi_0$ is a constant that can be set to {\it zero} by 
suitably choosing the scale $\mu$ and $r=\|\vec x\|$. Of course, the integration over 
$\vec y$ breaks the reciprocity symmetry. This also  
explains in what sense the reciprocity symmetry is realized in 
general relativity. It is further of interest to investigate the consequences of integrating Eq.~(\ref{eom:Phi:xy:1}) over the positions rather than the momenta.
Defining the momentum space `potential' by
$\phi_p(\vec p\,) \equiv \int d^2 x \Phi$,
and assuming the surface term 
$[2\pi r (d/dr)\Phi]_{r\rightarrow \infty}$ 
($r=\|\vec x\,\|$) vanishes,
one obtains the momentum space Laplace equation
\begin{equation}
  \nabla_{\vec p}^2\, \phi_p(\vec p\,) = \frac{2\kappa}{3} M \delta^2(\vec p\,)
\,,
\label{Laplace:GR limit:p}
\end{equation}
with the solution
\begin{equation}
  \phi_p = \phi_0^p + G_NM\ln\bigg(\frac{p^2}{\mu_p^2}\bigg)
 \,,\qquad (\mu_p=(c^3/G_N)\mu\,,\; p =\|\vec p\,\|)
\,.
\label{potential:Newton:p}
\end{equation}
Comparing Eqs.~(\ref{Laplace:GR limit:x}) and~(\ref{Laplace:GR limit:p})
and the corresponding solutions~(\ref{potential:Newton:x})
and~(\ref{potential:Newton:p}), one can observe how Max Born's original idea
is realized in linearized Hermitian Gravity. According to Max Born,
reciprocity symmetry should be implemented in the (quantum) theory of gravity
which states the dynamical equations of gravity (and thus also the solutions) 
should be invariant under the {\it reciprocity 
transformations} ({\it cf.} Eqs.~(\ref{reciprocity symmetry:tangent}) and~(\ref{J:dX-dp}))
\begin{equation}
  \vec x\rightarrow \frac{G_N}{c^3} \vec p
  \,,
  \qquad 
\vec p\rightarrow -\frac{c^3}{G_N} \vec x
\,.
\label{Born's reciprocity}
\end{equation}
In fact, the implementation of the reciprocity symmetry 
harnesses a larger symmetry than originally presupposed by Born.
This concept is illustrated by solving
the Laplace equation~(\ref{eom:Phi:2}), 
which does not respect the radial-angular split of 
Eqs.~(\ref{nabla:rad:3}--\ref{nabla:ang:3}). 
Assuming the solution does not depend on any of the angular components,
only the radial part of 
the Laplace operator~(\ref{Laplace:2d}) survives
\begin{eqnarray}
\bigg(\frac{d^2}{d R^2}
          + \frac{3}{R}\frac{d}{d R}\bigg)\Phi(R)
     = \frac{1}{R^3}\frac{d}{dR}R^3\frac{d}{dR}\Phi(R)
       = \frac{2\kappa}{3} M \delta^2(\vec z\,)\delta^2(\vec {\bar z}\,)
\,.\nonumber
\label{eom:Phi:3:naive}
\end{eqnarray}
The general solution outside of the source, $R\neq 0$, is of the form
\begin{eqnarray}
  \Phi(R) = \Phi_0 + \frac{\Phi_1}{R^2}
\,,\nonumber
\label{eom:Phi:4:naive}
\end{eqnarray}
where $\Phi_0$ is a physically unimportant, constant potential 
which can be set to zero, and $\Phi_1$ is the second constant of integration.
This constant can be determined by making use of the Gauss theorem
\begin{eqnarray}
  \int d^2 z d^2 {\bar z}\,\nabla^2\Phi 
  &=& \int_0^{2\pi}d\alpha \int_0^{2\pi}d\theta 
  \int_0^{2\pi}d\phi \frac{|\sin(2\phi)|}{2} 
  \int_0^\infty dR R^3 \frac{1}{R^3}\frac{d}{dR}\left(R^3\frac{d}{dR}\right)\Phi
 \nonumber\\
   &=& 8\pi^2 \bigg[R^3\frac{d}{dR}\frac{\phi_1}{R^2}\bigg]_{R\rightarrow \infty}
    = -16\pi^2\phi_1   
         = \frac{2\kappa}{3} M
\,,
\label{eom:Phi:5:naive}
\end{eqnarray}
such that the integration constant is fixed at $\phi_1 = -\frac{\kappa M}{24 \pi^2}$.
The constant $\kappa$ can be fixed by requiring that 
the correct Newtonian limit (Poisson equation)
is recovered in the three dimensional 
case (see section~\ref{Three-Dimensional Case});
the result is $\kappa=48\pi G_N$.
The factor $8\pi^2$ in Eq.~(\ref{eom:Phi:5:naive}) 
comes from the integration over $U(2)$ whose unit
volume is ${\rm Vol}[U(2)]=8\pi^2$, which is to be compared with the volume
of $SO(2)$ occuring in GR, ${\rm Vol}[SO(2)]=2\pi$. 
Thus, the gauge invariant potential for a static point-like mass is given by
\begin{equation}
 \Phi = -\frac{\kappa M}{24(\pi R)^2}  = -\frac{\kappa M}{48 \pi^2 z \bar z}
       = -\frac{\kappa M}{24\left(\pi\right)^2\big(r^2 + (G_Np/c^3)^2\big)}
\,.
\label{eom:Phi:6:naive}
\end{equation}
First, notice this phase space potential differs from
the solution~(\ref{potential:Newton:x}) of general relativity. Yet, in contrast to  
the Newton potential given by~(\ref{potential:Newton:x}), 
the Hermitian Gravity potential~(\ref{eom:Phi:6:naive}) 
respects the reciprocity symmetry~(\ref{Born's reciprocity}).
In fact, the discrete symmetry~(\ref{Born's reciprocity})
 is promoted here to a $U(1)$ symmetry:
\begin{equation}
  z\rightarrow z {\rm e}^{\imath \alpha}
  \,,\qquad
  \bar z \rightarrow \bar z {\rm e}^{-\imath \alpha}
\qquad (\alpha\in[0,2\pi))
\,.
\label{reciprocity:HG}
\end{equation}
It is not known whether the solutions of 
Eqs.~(\ref{nabla:rad:3}) exhibit the same enhanced symmetry.

 As discussed above, the Hermitian Gravity potential~(\ref{eom:Phi:6:naive}) 
 is in fact a phase space 
 potential that has to be integrated over to reach the general relativistic limit. Performing this integration for equation
 (\ref{eom:Phi:6:naive}):

\begin{eqnarray}
 \phi &=& \int d^2 y \Phi
                = -\frac{\kappa M}{12\pi} \int_0^Y \frac{r_y dr_y}{r^2 + r_y^2}
\nonumber\\
              &=& -\frac{\kappa M}{24\pi}\ln(r^2+r_y^2)|_{r_y=0}^{r_y=Y}
               = \frac{\kappa M}{24 \pi}\ln\Big(\frac{r^2}{r^2+Y^2}\Big)
\,,
\label{eom:Phi:7:naive}
\end{eqnarray}
where $Y$ has been added as an ultraviolet regulator. The result~(\ref{eom:Phi:7:naive})
diverges logarithmically as $Y$ tends to infinity. The solution does not satisfy the differential equation (\ref{Laplace:GR limit:x}) for finite values of $Y$. This, in fact, does not pose a problem, because one can reintroduce the integration constant $\Phi_0$ which was previously chosen to be zero such that the full solution reads
\begin{eqnarray}
\phi = \frac{\kappa M}{24 \pi}\ln\Big(\frac{r^2}{r^2+Y^2}\Big)+2 \pi\Phi_0 Y^2,\nonumber
\end{eqnarray}
and by giving $\Phi_0$ an appropriate, infinitesimally small value determined by the equation
\begin{eqnarray}
\left(\frac{-\kappa M}{24\pi}\ln(r^2 + Y^2) + 2\pi \Phi_0 Y^2 \right) \Bigg|_{Y \rightarrow \infty} = \frac{-\kappa M}{24 \pi}\ln(\mu^2).\nonumber
\end{eqnarray}
One obtains the regular limit as $Y$ goes to infinity
\begin{equation}\label{solmeth2}
\phi = \frac{\kappa M}{24 \pi} \ln(\frac{r^2}{\mu^2}),
\end{equation}
which for the suitable 
value~\footnote{According to the analysis of section~\ref{Three-Dimensional Case}
this is not the correct value of $\kappa$.} of $\kappa\rightarrow 24 \pi G_N$
reproduces (\ref{potential:Newton:x}), 
the Newtonian limit of general relativity
\begin{equation}
  \phi_{GR}(r) \rightarrow 2G_NM\ln\Big(\frac{r}{\mu}\Big)
\,.
\label{eom:Phi:9:naive}
\end{equation}
Thus, it has been shown that integrating out the momenta and solving the Poisson equation give completely equivalent results, regardless of the order the operations are performed in. In both cases, the momentum `radius' was taken to the infinite limit. In deriving (\ref{solmeth2}), it was necessary for the result to be a proper solution of the Poisson equation, while in deriving (\ref{potential:Newton:x}) the infinite limit was used to omit the surface term. \\
The linearized theory result is, of course,
incorrect. In full general relativity a point mass in two spatial dimensions sources
a conical singularity at the particle location, but no
gravitational field away from the particle. It is of interest to
find out the gravitational potential in
full Hermitian Gravity. This will be left for a future publication;
the three-dimensional case of linearized Hermitian Gravity will now be addressed.

\subsection{Three-Dimensional Case}
\label{Three-Dimensional Case}

The three-dimensional case, $d$=3, is in many ways similar to the two-dimensional case. The three-dimensional complex space \{$\vec{z},\vec{\bar{z}}$\} can be spanned by six real dimensions \{$R,\theta,\phi,\alpha,\beta,\gamma$\} representing a polar coordinate basis. In terms of this basis, the Cartesian coordinates
can be written as
\begin{eqnarray}
z^1=\frac{R}{\sqrt{2}}{\rm e}^{\imath \alpha} {\rm e}^{\imath \beta}\sin{(\phi)}\sin{(\theta)},&&\qquad z^{\bar{1}}=\frac{R}{\sqrt{2}}{\rm e}^{-\imath \alpha} {\rm e}^{-\imath \beta}\sin{(\phi)}\sin{(\theta)}\nonumber\\
z^2=\frac{R}{\sqrt{2}}{\rm e}^{\imath \alpha} \cos{(\phi)}\sin{(\theta)},&&\qquad z^{\bar{2}}=\frac{R}{\sqrt{2}}{\rm e}^{- \imath \alpha} \cos{(\phi)}\sin{(\theta)}\nonumber\\
z^3=\frac{R}{\sqrt{2}}{\rm e}^{\imath \alpha}{\rm e}^{\imath \gamma} \cos{(\theta)},&&\qquad z^{\bar{3}}=\frac{R}{\sqrt{2}}{\rm e}^{-\imath \alpha}{\rm e}^{-\imath \gamma} \cos{(\theta)}\nonumber
\end{eqnarray}
In this case, the Laplacian will be calculated using (\ref{fastlaplace}). To do so, the (spatial part of the) metric needs to be derived first. The infinitesimal transformations corresponding to the above coordinates are given by:
\begin{eqnarray}
{\rm d}z^1&=&\frac{1}{\sqrt{2}}\Big({\rm e}^{\imath \alpha} {\rm e}^{\imath \beta}\sin{(\phi)}\sin{(\theta)}{\rm d}R + R{\rm e}^{\imath \alpha}{\rm e}^{\imath \beta} \cos{(\phi)}\sin{(\theta)}{\rm d}\phi+ R{\rm e}^{\imath \alpha} {\rm e}^{\imath \beta}\sin{(\phi)}\cos{(\theta)}{\rm d}\theta\nonumber\\
&&+ \imath R {\rm e}^{\imath \alpha} {\rm e}^{\imath \beta}\sin{(\phi)}\sin{(\theta)}{\rm d}\alpha+\imath R {\rm e}^{\imath \alpha} {\rm e}^{\imath \beta}\sin{(\phi)}\sin{(\theta)}{\rm d}\beta\Big)\nonumber,\\
{\rm d}z^2&=&\frac{1}{\sqrt{2}}\Big({\rm e}^{\imath \alpha} \cos{(\phi)}\sin{(\theta)}{\rm d}R - R{\rm e}^{\imath \alpha} \sin{(\phi)}\sin{(\theta)}{\rm d}\phi\nonumber\\
&&+ R{\rm e}^{\imath \alpha} \cos{(\phi)}\cos{(\theta)}{\rm d}\theta + \imath R {\rm e}^{\imath \alpha} \cos{(\phi)}\sin{(\theta)}{\rm d}\alpha\Big)\nonumber,\\
{\rm d}z^3&=&\frac{1}{\sqrt{2}}\Big({\rm e}^{\imath \alpha} {\rm e}^{\imath \gamma}\cos{(\theta)}{\rm d}R - R{\rm e}^{\imath \alpha} {\rm e}^{\imath \gamma}\sin{(\theta)}{\rm d}\theta\nonumber\\
&&+ \imath R{\rm e}^{\imath \alpha} {\rm e}^{\imath \gamma}\cos{(\theta)}{\rm d}\alpha + \imath R {\rm e}^{\imath \alpha} {\rm e}^{\imath \gamma}\cos{(\theta)}{\rm d}\gamma\Big),\nonumber
\end{eqnarray}
and their complex conjugates. The above are inserted 
into the metric in Cartesian coordinates,
${\rm ds}^2 = 2\left({\rm d}z^1{\rm d}z^{\bar{1}}+{\rm d}z^2{\rm d}z^{\bar{2}}+{\rm d}z^3{\rm d}z^{\bar{3}}\right)$,  to obtain the metric in polar coordinates
\begin{eqnarray}{\label{3dpolmetric}}
{\rm ds}^2 &=& {\rm d}R^2 + R^2 \sin^2{(\theta)} {\rm d}\phi^2 + R^2 {\rm d}\theta^2 + R^2{\rm d}\alpha^2 + R^2 \sin^2{(\phi)} \sin^2{(\theta)} {\rm d}\beta^2 \nonumber\\
&& + R^2 \cos^2{(\theta)} {\rm d}\gamma^2 + 2 R^2 \sin^2{(\phi)} \sin^2{(\theta}) {\rm d}\alpha {\rm d}\beta + 2 R^2 \cos^2{(\theta)} {\rm d}\alpha {\rm d}\gamma,\nonumber
\end{eqnarray}
or, equivalently, in matrix notation
\begin{eqnarray}
C_{i\bar{j}}=\left(\begin{array}{cccccc}
1&0&0&0&0&0\\
0&R^2&0&0&0&0\\
0&0&R^2 \sin^2{(\theta)}&0&0&0\\
0&0&0&R^2&R^2 \sin^2{(\phi)} \sin^2{(\theta)}&R^2 \cos^2{(\theta)}\\
0&0&0&R^2 \sin^2{(\phi)} \sin^2{(\theta)}&R^2 \sin^2{(\phi)} \sin^2{(\theta)}&0\\
0&0&0&R^2 \cos^2{(\theta)}&0&R^2 \cos^2{(\theta)}\end{array}
\right).\nonumber
\end{eqnarray}
From this, the quantities necessary for calculating the Laplacian can straightforwardly be found. The determinant $C$ is found to be
\begin{eqnarray}
C=R^{10} \left( \cos^2{(\phi)} \sin^2{(\phi)} \cos^2{(\theta)} \sin^6{(\theta)}\right),\nonumber
\end{eqnarray}
while the inverse of the spatial metric yields
\begin{eqnarray}
C^{i\bar{j}}=\left(
\begin{array}{cccccc}
1&0&0&0&0&0\\
0&\frac{1}{R^2}&0&0&0&0\\
0&0&\frac{1}{R^2 \sin^2{(\theta)}}&0&0&0\\
0&0&0&\frac{1}{R^2 \cos^2{(\phi)} \sin^2{(\theta)}}&\frac{-1}{R^2 \cos^2{(\phi)} \sin^2{(\theta)}}&\frac{-1}{R^2 \cos^2{(\phi)} \sin^2{(\theta)}}\\
0&0&0&\frac{-1}{R^2 \cos^2{(\phi)} \sin^2{(\theta)}}&\frac{4}{R^2 \sin^2{(2 \phi)} \sin^2{(\theta)}}&\frac{1}{R^2 \cos^2{(\phi)} \sin^2{(\theta)}}\\
0&0&0&\frac{-1}{R^2 \cos^2{(\phi)} \sin^2{(\theta)}}&\frac{1}{R^2 \cos^2{(\phi)} \sin^2{(\theta)}}&\frac{1}{R^2 \cos^2{(\phi)} \sin^2{(\theta)}}+\frac{1}{R^2\cos^2{(\theta)}}\end{array}\right).\nonumber
\end{eqnarray}
Combining the above results with (\ref{fastlaplace}), one readily finds the Laplacian to be
\begin{eqnarray}\label{3dlaplacian}
\nabla^2 &=& \frac{\partial^2}{\partial R^2}+\frac{5}{R}\frac{\partial}{\partial R}+\frac{1}{R^2}\frac{\partial^2}{\partial \theta^2}+\frac{1}{R^2}\left(3\cot{(\theta)}-\tan{(\theta)}\right)\frac{\partial}{\partial \theta}\nonumber\\
&&+\frac{1}{R^2 \sin^2{(\theta)}}\left(\frac{\partial^2}{\partial \phi^2} + 2\cot{(2\theta)}\frac{\partial}{\partial \phi}+\frac{4}{\sin^2{(2\phi)}}\frac{\partial^2}{\partial \beta^2}+\tan^2{(\theta)} \frac{\partial^2}{\partial \gamma^2}\right)\nonumber\\
&&+\frac{1}{R^2 \cos^2{(\phi)} \sin^2{(\theta)}}\left(\frac{\partial^2}{\partial \alpha^2} - 2 \frac{\partial}{\partial \alpha}\frac{\partial}{\partial \beta} - 2\frac{\partial}{\partial \alpha} \frac{\partial}{\partial \gamma}+2\frac{\partial}{\partial \beta}\frac{\partial}{\partial \gamma}+\frac{\partial^2}{\partial \gamma^2}\right).\nonumber
\end{eqnarray}
Proceeding along similar lines as in the two-dimensional case, the differential equation (\ref{eom:Phi:2}) generated by this Laplacian will be solved for a $d=3$ static point-like source on phase space
\begin{eqnarray}
\rho_{HG}=M \delta^3\left(\vec{z}\right)\delta^3\left(\vec{\bar{z}}\right).\nonumber
\end{eqnarray}
It is possible to split the three-dimensional Laplacian into two reciprocally invariant parts. The radial, reciprocally invariant piece stays the same as in (\ref{nabla:rad:3}), while (\ref{nabla:ang:3}) changes suitably to represent the entire Laplacian with (\ref{nabla:rad:3}) removed. These equations are even more laborious than before, and thus seem to be too difficult to solve at this moment. Instead, it is possible to solve the Laplacian for a potential which does not depend on the angular variables for which (\ref{eom:Phi:2}) simplifies to
\begin{equation}
\frac{1}{R^5}\frac{d}{dR}\left(R^5\frac{d}{dR}\right)\Phi(R) = \frac{2\kappa}{3} M \delta^3(\vec z\,)\delta^3(\vec {\bar z})
\,.
\label{Phieom3d}
\end{equation}
Away from the source, such that $R\neq 0$, the general solution is given by
\begin{eqnarray}
\Phi = \Phi_0 + \frac{\Phi_1}{R^4}.\nonumber
\end{eqnarray}
Once again, the constant $\Phi_0$ can be set to zero~\footnote{Unlike in the two-dimensional case, it will not be necessary to keep $\Phi_0$ for reasons of convergence.} and Gauss's theorem is applied to fix the integration constant $\Phi_1$ by considering an integration in (\ref{Phieom3d}) over the six-dimensional phase space 
\begin{eqnarray}\label{constfix3d}
\int d^3 z \int d^3 \bar{z}\,\nabla^2 \Phi &=&\int_0^{2\pi} d\alpha \int_0^{2\pi} d\beta \int_0^{2\pi} d\gamma \int_0^{2\pi}  d\phi  \int_0^{\pi}  d\theta \int_0^{\infty}  dR   \nonumber\\
&&\times \left|\cos{(\phi)} \sin{(\phi)} \cos{(\theta)} \sin^3{(\theta)}\right| \frac{d}{dR}\left(R^5 \frac{d}{dR}\right)\Phi(R)\nonumber\\
&=&8 \pi^3 \int_0^{\infty}  dR  \frac{d}{dR}\left(R^5 \frac{d}{dR}\right)\Phi(R) = 8 \pi^3 \left(R^5 \frac{d}{dR}\right)\Bigg|_{R\rightarrow \infty}\Phi(R)\nonumber\\
&=&-32 \pi^3 \Phi_1 = \frac{2\kappa}{3}\kappa M,\nonumber
\end{eqnarray}
where Vol[U(3)]$=8\pi^3$ has been used, upon which the constant $\Phi_1$ is found to be
\begin{eqnarray}
\Phi_1=\frac{-\kappa M}{48 \pi^3}.\nonumber
\end{eqnarray}
Thus the solution for the phase space potential with constants fixed is 
\begin{equation}
\Phi = \frac{-\kappa M}{48 \pi^3 R^4}
        =\frac{-\kappa M}{192 \pi^3 (z\bar{z})^2}.
\label{Phi:3d:HG}
\end{equation}
Just as in the two-dimensional case, this solution respects the $U(1)$ 
reciprocity symmetry~(\ref{reciprocity:HG}).
To obtain the potential of general relativity, 
the momenta are integrated 
over~\footnote{Here, the relations $R^2=\vec{x}^2+\vec{y}^2$, 
$r=||\vec{x}||$, and $r_y=||\vec{y}||$ are made use of.}
\begin{eqnarray}
\phi_{GR}&=&\int_0^{\infty} d^3 y \Phi = \int_0^{\infty} d r_y 4\pi r_y^2 \Phi(R) = -\int_0^{\infty} d r_y \frac{\kappa M r_y^2}{12 \pi^2 \left(r^2+r_y^2\right)}\nonumber\\ 
&=& \frac{-\kappa M}{48 \pi r},\nonumber
\end{eqnarray}
and one determines $\kappa=48 \pi G_N$ which results 
in the correct Newtonian limit of general relativity
\begin{eqnarray}
\phi_{GR}=-\frac{G_N M}{r}.\nonumber
\end{eqnarray}
Hence, it has been shown that Hermitian Gravity does indeed reproduce
the same Newtonian limit as general relativity, provided
$\kappa=48 \pi G_N$. In this case the potential~(\ref{Phi:3d:HG}) 
is completely fixed,
\begin{eqnarray}
\Phi = -\frac{G_N M}{4 \pi^2 (z\bar{z})^2}
\,,\qquad z\bar{z} = \frac{R^2}{2}
\,.\nonumber
\label{Phi:3d:HG:2}
\end{eqnarray}
Moreover, upon inserting $\kappa=48 \pi G_N$ into 
the two-dimensional potential~(\ref{solmeth2}), one obtains 
$\phi_{\rm GR}\rightarrow 4G_NM\ln(r/\mu)=2\phi_{GR}$,
which is a factor of two larger than the GR potential~(\ref{eom:Phi:9:naive}).
This could in principle be used as a test to distinguish between GR and HG,
since a straight cosmic string in Hermitian Gravity will generate
a deficit angle that is twice as large as that of general relativity.
 \\
Previously, there have been no known 
solutions~\footnote{According to Nakahara~\cite{Nakahara:2003nw}.
It should be noted however that Ref.~\cite{Nakahara:2003nw} 
defines Hermitian geometry by requiring that all covariant derivatives
acting on a vielbein vanish, 
$\nabla^\mu e_\nu^a = 0, \nabla^{\bar \mu} e_\nu^a = 0$.
These conditions then generate different connection coefficients
than the ones used here, 
which are consistent with the weaker conditions:
$\nabla^\mu C_{\alpha\bar\beta}=0$, 
 $\nabla^{\bar \mu} C_{\alpha\bar\beta}=0$~\cite{Mantz:2008hm}.} 
to a theory of Hermitian Gravity, but by going to 
the linearized theory the first such solution has been found.
 This opens perspectives for other linearized solutions, 
such as gravitational waves.

\section{Motion of Massive Particles and Light Deflection}

Having derived the solutions to the gauge invariant potentials in the previous section, it is now possible to compute the motion of massive particles and massless particles (light deflection). For the non-relativistic massive particle, one assumes that the particle moves slowly with respect to the speed of light, $\frac{d z^i}{d\tau}\ll \frac{d z^0}{d\tau}$ and  $\frac{d z^{\bar{i}}}{d\tau}\ll \frac{d z^{\bar{0}}}{d\tau}$, and the field acting on the particle is static such that $\partial_0 C_{mn} = 0$ and $\partial_{\bar{0}} C_{mn}=0$. In addition, we continue to work in the framework where the gravitational field is weak, which we have already used in linearizing the theory. The geodesic equation is given by
\begin{equation}\label{geodesiceq}
\mbox{\boldmath{$\frac{d^2 z^r}{d\tau^2}=\Gamma^r_{mn}\frac{d z^m}{d\tau}\frac{d z^n}{d\tau}$}},
\end{equation}
and the Christoffel symbols are given in equation (\ref{Christoffel2}). Under the above conditions, the various equations of motion obtained from the geodesic equation are
\begin{eqnarray}
\frac{d^2 z^0}{d\tau^2}=0,&&\qquad \frac{d^2 z^{\bar{0}}}{d\tau^2}=0,\nonumber\\
\frac{d^2 z^i}{d\tau^2}=\frac{d z^0}{d\tau}\frac{d z^{\bar{0}}}{d\tau}\partial_{\bar{i}}H_{\bar{0}0},&&\qquad \frac{d^2 z^{\bar{i}}}{d\tau^2}=\frac{d z^0}{d\tau}\frac{d z^{\bar{0}}}{d\tau}\partial_i H_{\bar{0}0}.\nonumber
\end{eqnarray}
Using the metric (\ref{bardpotmetric}) and the condition~\footnote{In the limit $dy^{\mu}\rightarrow 0$, this properly corresponds to the condition in general relativity $g_{\mu\nu}\frac{dx^{\mu}}{d\tau}\frac{dx^{\nu}}{d\tau}=-1$.} $C_{mn}\frac{dz^m}{d\tau}\frac{dz^n}{d\tau}=-1$, which in this case simplifies to $\frac{dz^{0}}{d\tau}\frac{dz_{0}}{d\tau}=-\frac{1}{2}$, the above equations become
\begin{eqnarray}
\frac{d^2 z^0}{d\tau^2}=0,&&\qquad \frac{d^2 z^{\bar{0}}}{d\tau^2}=0,\nonumber\\
\frac{d^2 z^i}{d\tau^2}=-\partial_{\bar{i}}\Phi,&&\qquad \frac{d^2 z^{\bar{0}}}{d\tau^2}=-\partial_i \Phi.\nonumber
\end{eqnarray}
To obtain the general relativistic limit for the spatial equations of motion, it is necessary to integrate the phase space potential $\Phi$ over the momenta. Introducing this integration, the equation for $z^i$ can also be written in terms of its real and imaginary parts by going to the $x-y$ basis
\begin{equation}
\frac{d^2 x^i}{d\tau^2}+\imath \frac{d^2 y^i}{d\tau^2}=-\int d^3 y \,\left(\partial_{x^i}-\imath \partial_{y^i}\right)\Phi. \label{geomot1}
\end{equation}
The imaginary part of the equation, $\imath \frac{d^2 y^i}{d\tau^2}=\int d^3 y \,\imath \partial_{y^i} \Phi$, represents a boundary term which can be 
neglected~\footnote{To see this, one can consider the integral in polar coordinates, neglecting the derivative temporarily
\[
\int_0^\infty d r_y \,  4\pi r_y^2 \Phi=-\int_0^\infty d r_y \,  \frac{G_N M r_y^2}{\pi \left(x^2 + r_y^2\right)^2} =  \frac{G_N M\left(\frac{r_y}{2\left(x^2+r_y^2\right)}+\frac{\arctan\left(\frac{r_y}{x}\right)}{2x}\right)}{\pi}\Bigg|^{r_y=\infty}_{r_y=0}.
\]
The first term drops out in both limits while the second term leaves $\frac{G_N M}{4x}$, where we have made use of $\arctan(\infty)=\frac{\pi}{2}$. Acting on this with the $\partial_y$ derivative that we have previously ignored, we find as a result $\frac{d^2 y^i_{GR}}{d\tau^2}=0$.
}.  On the other hand, the real part of equation (\ref{geomot1}) gives
\[
\frac{d^2 x^i}{d\tau^2}=-\int d^3 y \,\partial_{x} \Phi. 
\]
Previously, the relation between the phase space potential and the general relativistic potential in the three-dimensional case was found: $\phi_{GR}=\int d^3 y \, \Phi$. Applying this, the above equation simplifies to
\[ 
\frac{d^2 x^i_{GR}}{d\tau^2}=-\partial_{x} \phi_{GR},
\]
which is exactly the acceleration equation for a massive particle in the non-relativistic static limit of general relativity~\footnote{The subscript $GR$ has been used to accentuate this correspondence.}. Hence, the correct result has been reproduced in Hermitian Gravity. \\
For light deflection, we work within the same framework of weak and time-independent fields, but no longer is the assumption of non-relativistic velocities valid. It can be shown for time-independent fields from the decomposition of the Bardeen potentials in section 3 that the dynamics are generated by
\begin{equation}\label{eomlight}
\nabla^2 \phi = \frac{2\kappa \rho}{3},\qquad\nabla^2 \psi = \frac{\kappa\rho}{3},\qquad E=0,\qquad B=0.
\end{equation}
Thus, the metric takes on the form $H_{\mu\bar{\nu}}={\rm diag}\left(-2\phi, -2\psi, -2\psi, -2\psi\right)$. For this metric, the geodesic equation (\ref{geodesiceq}) generates the following accelerations
\begin{eqnarray}\label{ldsol1}
\frac{d^2 z^0}{d\tau^2}&=&-2\Gamma^0_{0i}\frac{d z^0}{d\tau}\frac{d z^i}{d\tau}-2\Gamma^0_{0\bar{i}}\frac{d z^0}{d\tau}\frac{d z^{\bar{i}}}{d\tau}
=-4\frac{d z^0}{d\tau}\frac{d z^i}{d\tau}\partial_i \phi,\nonumber\\
\frac{d^2 z^i}{d\tau^2}&=&-\Gamma^i_{ja}\frac{d z^j}{d\tau}\frac{d z^a}{d\tau}-2\Gamma^i_{0\bar{0}}\frac{d z^0}{d\tau}\frac{d z^{\bar{0}}}{d\tau}-2\Gamma^i_{\bar{j}a}\frac{d z^{\bar{j}}}{d\tau}\frac{d z^a}{d\tau}=4\frac{d z^i}{d\tau}\frac{d z^j}{d\tau}\partial_j \psi - 2 \frac{d z^0}{d\tau}\frac{d z^{\bar{0}}}{d\tau}\partial_{\bar{i}}\phi-2\frac{d z^a}{d\tau}\frac{d z^{\bar{a}}}{d\tau}\partial_{\bar{i}}\psi,\nonumber\\
\end{eqnarray}
and their complex conjugates. From the line element $C_{\mu\bar{\nu}}\frac{dz^{\mu}}{d\tau}\frac{dz^{\bar{\nu}}}{d\tau}=0$, one can conclude $\frac{dz^{\mu}}{d\tau}\frac{dz_{\mu}}{d\tau}=0$ and thus $\frac{dz^{0}}{d\tau}\frac{dz_{0}}{d\tau}=\frac{dz^{i}}{d\tau}\frac{dz_{i}}{d\tau}$ at zeroth order~\footnote{It is valid to approximate this relation to zeroth order, since in equation (\ref{ldsol1}) the indices are raised and lowered with the Minkowski metric. Using the full metric would generate additional second-order correction terms which are negligible.} upon which the second equation above reduces to
\[
\frac{d^2 z^i}{d\tau^2}=4\frac{d z^i}{d\tau}\frac{d z^j}{d\tau}\partial_j \psi - 2 \frac{d z^j}{d\tau}\frac{d z_j}{d\tau}\partial_{\bar{i}}\phi-2\frac{d z^j}{d\tau}\frac{d z_j}{d\tau}\partial_{\bar{i}}\psi.
\]
These are to be compared with the accelerations found in general relativity
\begin{eqnarray}
\frac{d^2 x^0_{GR}}{d\tau^2}&=&-2 \frac{dx_{GR}^0}{d\tau}\frac{dx_{GR}^i}{d\tau}\partial_i \phi_{GR}\label{GRtime},\\
\frac{d^2 x^i_{GR}}{d\tau^2}&=&2 \frac{dx_{GR}^i}{d\tau}\frac{dx_{GR}^j}{d\tau}\partial_j \phi_{GR}-2 \eta_{jk}\frac{dx_{GR}^j}{d\tau}\frac{dx_{GR}^k}{d\tau} \partial_i \phi_{GR}.\label{GRspace}
\end{eqnarray}
%
%
%
As before, it is beneficial to rewrite the equations in Hermitian Gravity in the $x-y$ basis. The acceleration of the time direction then reads~\footnote{The relevant transformations here are $z^{\mu}=\frac{1}{\sqrt{2}}\left(x^{\mu}+iy^{\mu}\right)$ and $\frac{\partial}{\partial z^{\mu}}=\frac{1}{\sqrt{2}}\left(\frac{\partial}{\partial x^{\mu}}-i\frac{\partial}{\partial y^{\mu}}\right)$.} 
\begin{eqnarray}
\frac{d^2 x^0}{d\tau^2}+i\frac{d^2 y^0}{d\tau^2}&=&-2\frac{d x^0 + i d y^0}{d\tau}\frac{d x^i + i d y^i}{d\tau}\left(\partial_{x^i}-i\partial_{y^i}\right) \phi.\nonumber
\end{eqnarray}
This expression is more complicated than the corresponding one in general relativity. However, if one considers the limit for sub-Planckian scales, where the momenta $y^{\mu}\approx 0$ are negligible and very slowly changing such that $\frac{d y^{\mu}}{d\tau}=0$, then one obtains the result (\ref{GRtime}) from general relativity~\footnote{In the massive case, the $y$-acceleration$\frac{d^2 y^{\mu}}{d\tau^2}$ was found to be zero as a result of solving the integration (\ref{geomot1}). It is not as straightforward to show that this is true for the massless case, however we will show that the above conditions for sub-Planckian scales will lead to consistent results.} 
\begin{equation}\label{zerogeoeq}
\frac{d^2 x^0}{d\tau^2}=-2\frac{dx^0}{d\tau}\frac{dx^i}{d\tau}\partial_{x^i}\phi_{GR}.
\end{equation}
For the spatial geodesic equation, the same coordinate change generates
\begin{eqnarray}
\frac{d^2 x^i}{d\tau^2}+i\frac{d^2 y^i}{d\tau^2}&=&2\frac{d x^i+i d y^i}{d\tau}\frac{d x^j+i d y^j}{d\tau}\left(\partial_{x^j}-i\partial_{y^j}\right) \psi -  \frac{d x^j+i dy^j}{d\tau}\frac{d x_j+i dy_j}{d\tau}\left(\partial_{x^i}+i\partial_{y^i}\right)\phi\nonumber\\
&&-\frac{d x^j+i dy^j}{d\tau}\frac{d x_j+i dy_j}{d\tau}\left(\partial_{x^i}+i\partial_{y^i}\right)\psi.\nonumber
\end{eqnarray}
Taking once again the sub-Planckian limit and performing an integration over the allowed momenta to obtain the GR potentials
\begin{equation}\label{spacegeoeq}
\frac{d^2 x^i}{d\tau^2}=2\frac{dx^i}{d\tau}\frac{dx^j}{d\tau}\partial_{x^j}\psi_{GR}-\frac{dx^j}{d\tau}\frac{dx_j}{d\tau}\partial_{x^i}\phi_{GR}-\frac{dx^j}{d\tau}\frac{dx_j}{d\tau}\partial_{x^i}\psi_{GR},
\end{equation}
where we observe that this result is actually quite different than in GR. If one takes $\phi=\psi$, then indeed the general relativistic limit (\ref{GRspace}) is obtained~\footnote{The quantity on the right hand side of equation (\ref{GRspace}) is actually nothing more than double the transverse gradient, the total gradient of the particle's motion with the longitudinal component removed. As will be shown, it describes how a photon passing by a massive source has its path altered orthogonally to the direction it is moving in such that it bends. The overall factor of two is a significant factor, which has been experimentally verified scrutinously, and distinguishes Einstein's general relativity from Newton's theory (where the overall factor is not present).}. However, from the equations of motion (\ref{eomlight}), it can be concluded for any local source that $\phi=2\psi$ in the case of Hermitian Gravity. The natural way to proceed then will be to see by what magnitude light is deflected in this theory.\\
To begin, we consider the equation of motion for the zero component (\ref{zerogeoeq}). Introducing the notation $\dot{x}^0=\frac{dx^0}{d\tau}$, a simple rewriting yields
\begin{eqnarray}
\frac{\ddot{x}^0}{\dot{x}^0}=-2\frac{dx^i}{d\tau}\partial_{x^i}\phi_{GR}.\nonumber
\end{eqnarray}
Integrating both sides with respect to $\tau$ leads to
\begin{equation}\label{deltaphi}
\ln{\left(\frac{\dot{x}^0}{v^0}\right)}=-2\int \frac{dx^i}{d\tau}\partial_{x^i}\phi_{GR}\,d\tau = -2\int\nabla\phi_{GR}\cdot d\vec{x}=-2\Delta\phi_{GR}.
\end{equation}
Here the quantity $\Delta \phi_{GR}$ is just the change in $\phi_{GR}$ over the region of integration and $v^0$ is the initial velocity of the time direction. If we consider some massive source at $(0,0,0)$ in coordinates $(x^{||},x^{\perp},z)$~\footnote{The coordinates $x^{\perp}$, $x^{||}$, and $z$ form an orthonormal basis in three-dimensional space.}, then at any spatial point infinitely far away from this source the potential $\phi_{GR}$ vanishes, i.e. $\phi_{GR}(\infty)=0$. Hence, if the integration in (\ref{deltaphi}) is chosen to be from $-\infty$ to $\infty$, the result $\Delta \phi_{GR}$ will be null. Analogously if instead the integration is fixed from $-\infty$ to some point $a$, then $\Delta \phi_{GR}$ will be exactly the value of $\phi_{GR}(a)$. As a result, we can (re)introduce the function $\phi_{GR}(x)$:
\begin{eqnarray}
\ln{\left(\frac{\dot{x}^0}{v^0}\right)}=-2\int_{-\infty}^{x}\nabla\phi_{GR}\cdot d\vec{x}=-2\phi_{GR}(x).\nonumber
\end{eqnarray}
Exponentiating both sides of the equation produces
\begin{eqnarray}
\dot{x}^0=\exp\left(-2\phi_{GR}(x)\right)\approx 1-2\phi_{GR}(x).\nonumber
\end{eqnarray}
The above equation implies that, for an observer at infinity, a particle travelling from infinity to the source will experience time dilation. Since $\dot{x}^0$ is unity at both $x=-\infty$ and $x=\infty$, it is a conserved quantity along this path. We consider now that this massless particle begins at point $(b,-\infty,0)$, travelling strictly along the $x^{||}$ axis towards the point $(b,0,0)$. In the absence of the source, the particle would continue travelling purely along the $x^{||}$ axis. However, with the source present, the $x^{\perp}$ coordinate is also affected~\footnote{The $z$ coordinate is left unaffected, since both the source and the particle are at $z=0$. In essence, the motion is just two-dimensional.}. For this scenario, the geodesic equations for the spatial components (\ref{spacegeoeq}) can be expressed as
\begin{eqnarray}\label{spacegeoeq2}
\ddot{x}^{\perp}&=&2\dot{x}^{\perp}\left(\dot{x}^{\perp}\partial_{\perp}\psi_{GR}+\dot{x}^{||}\partial_{||}\psi_{GR}\right)-\left((\dot{x}^{\perp})^2+(\dot{x}^{||})^2\right)\left(\partial_{\perp}\phi_{GR}+\partial_{\perp}\psi_{GR}\right),\nonumber\\
\ddot{x}^{||}&=&2\dot{x}^{||}\left(\dot{x}^{\perp}\partial_{\perp}\psi_{GR}+\dot{x}^{||}\partial_{||}\psi_{GR}\right)-\left((\dot{x}^{\perp})^2+(\dot{x}^{||})^2\right)\left(\partial_{||}\phi_{GR}+\partial_{||}\psi_{GR}\right),
\end{eqnarray}
where $\partial_{\perp}=\frac{\partial}{\partial x^{\perp}}$ and $\partial_{||}=\frac{\partial}{\partial x^{||}}$. Combining the relations $\dot{x}^0=1-2\phi_{GR}$ and $C_{\mu\bar{\nu}}\frac{dz^{\mu}}{d\tau}\frac{dz^{\bar{\nu}}}{d\tau}=0$, we find up to first order in corrections
\begin{equation}\label{ydotrel}
(\dot{x}^0)^2=(\dot{x}^{\perp})^2+(\dot{x}^{||})^2 = 1-4\phi_{GR}.
\end{equation}
Because we want to compare this to the light deflection generated by the Sun, we know the deflection angle will be very small. This angle $\alpha(x)$ is the tangent of the particle's transverse and parallel velocities, $\dot{x}^{\perp}$ and $\dot{x}^{||}$ respectively, so that $\alpha(x)=\tan\left(\frac{\dot{x}^{\perp}}{\dot{x}^{||}}\right)$. For very small angles, this can be well approximated as $\alpha(x)=\frac{\dot{x}^{\perp}}{\dot{x}^{||}}$. By manipulating (\ref{ydotrel}) with the use of $\alpha$, it can be shown that $\dot{x}^{||}$ varies due to the source's presence up to first order
\begin{eqnarray}
(\alpha^2+1)(\dot{x}^{||})^2 = 1-4\phi_{GR}\Rightarrow \dot{x}^{||}=1+O(\phi_{GR})+O(\alpha^2),\nonumber
\end{eqnarray}
where the $O(\alpha^2)$ term is negligible. It also implies, in conjunction with the definition of $\alpha$ for small deflection angles, that $\dot{x}^{\perp}\sim\alpha$.
With the above, it is now possible to examine the geodesic equations (\ref{spacegeoeq2}) by virtue of orders of magnitude. As has been the case throughout all of linearized Hermitian Gravity, only terms up to first order in corrections are kept~\footnote{Any term of the form $O(\alpha^2)$, $O(\alpha \phi_{GR})$, or $O(\phi_{GR}^2)$ is negligible.}. Integrating the second of the pair of equations provides a statement about the change in the parallel velocity $\Delta \dot{x}^{||}$:
\begin{eqnarray}
\Delta \dot{x}^{||}&=&2\int \left(1+O(\phi_{GR})\right)\alpha(x)\partial_{\perp}\psi_{GR}\,d\tau + 2\int \left(1+O(\phi_{GR})\right)\partial_{||}\phi_{GR}\cdot dx^{||}\nonumber\\
&&-\int (1+O(\phi_{GR}))\partial_{||}\left(\phi_{GR}+\psi_{GR}\right)d\tau.\nonumber
\end{eqnarray}
The first term on the right hand side vanishes by virtue of the fact that both $\psi_{GR}$ and $\alpha$ are small, and thus it is second order in corrections. In the second term, one can immediately discard the $O(\phi_{GR})$ term and integrate the remainder (it is just an integral of a derivative), $2\int \partial_{||}\phi_{GR}\cdot dx^{||}=2\Delta \psi_{GR}$. As described earlier, the change of $\phi_{GR}$, and so also $\psi_{GR}$, over the path the particle follows from negative infinity to infinity is zero. Thus the second term is also negligible. Finally, we multiply the third term by an overall factor of $\frac{\dot{x}^{||}}{\dot{x}^{||}}= \frac{\dot{x}^{||}}{1+O(\phi_{GR})}$. The numerator in this factor allows us to also rewrite this last term's lowest (first) order component as the integral of a derivative. Consequently it also disappears and we find that up to first order in corrections $\Delta \dot{x}^{||}=0$. This result is equivalent to the result in general relativity. If there was indeed some first order correction to this result, it would be measurable through the apparent redshifting of the photons while they pass the Sun. \\
The first of the equations (\ref{spacegeoeq2}) can be treated the same way. We observe that the first two terms are both of order $\alpha^2$ or higher due to the common factor of $\dot{x}^{\perp}$ and the potential $\psi_{GR}$ inside the parentheses. The third term vanishes for the same reason except that it is of order $\alpha^3$, since the distributive factor is $(\dot{x}^{\perp})^2$. Thus the final term remains in which we ignore the order $\phi_{GR}$ contribution from $(\dot{x}^{||})^2$, since it is negligible together with the fields. The only first order contribution is given by
\begin{eqnarray}
\dot{x}^{\perp}=-\int \partial_{\perp}\left(\phi_{GR}+\psi_{GR}\right)\,d\tau\nonumber,
\end{eqnarray}
which upon application of the relation $\alpha(x)=\frac{\dot{x}^{\perp}}{\dot{x}^{||}}$, leads to the desired result
\begin{eqnarray}
\alpha(x)\approx -\int\partial_{\perp}\left(\phi_{GR}+\psi_{GR}\right)\,d\tau.\nonumber
\end{eqnarray}
In the case of general relativity, one has $\phi_{GR}=\psi_{GR}$ and consequently $\alpha_{GR} \approx -2\int \partial_{\perp}\phi_{GR} \, d\tau$. However, for Hermitian Gravity the potentials obey a different relationship $\phi_{GR}=2\psi_{GR}$ and the resultant deflection angle 
\begin{eqnarray}
\alpha_{HG}\approx -\frac{3}{2}\int \partial_{\perp} \phi_{GR} \, d\tau = \frac{3}{4}\alpha_{GR},\nonumber
\end{eqnarray}
finds itself exactly between the predictions of the Newtonian theory
and general relativity. This contradicts the experimentally justified factor 
of two. The early experiments starting with Eddington's expedition in 1919, 
of which the merit is a continuing controversy, were quite imprecise. 
The factor $\frac{3}{4}$ found here could have been inside the margin of
error for these experiments. The more precise experiments from the late 
1960's onwards clearly rule out such a possibility though as the value
has been determined to be $\gamma=1.0002$ with an estimated standard error
of $0.002$~\cite{Robertson:1991}. The factor $\gamma$ is one that is
conventionally chosen for light deflection studies (it is one of the
ten dimensionless constants in the PPN formalism~\cite{Will:2005va}) 
and its value for 
general relativity is $\gamma=1$ (in Newton's theory, $\gamma$=0). These modern
experiments were performed on Earth using VLBI
(very-long-baseline-interferometry), as well as in space with the use of the
satellites Viking, Hipparcos, and more recently Cassini.

We have found that the theory correctly predicts the Newtonian limit and the nonrelativistic acceleration equation. However, it does not properly account for light deflection. The fundamental root of this problem is the relationship between the gauge invariant fields $\Phi=2\Psi$. The factor of two is a result of the dimensionality of the theory, and it produces the incorrect factor for the deflection angle. Since other relativistic corrections (Post-Newtonian) will similarly depend on $\Psi$, it is to be expected that they will also differ from general relativity. It is not yet known whether there is a method to fix this issue.  

One possible resolution lies in a different method of complexifying general 
relativity. As has been observed in a previous paper~\cite{Mantz:2008hm}, 
Hermitian Gravity displays favorable properties around singularities 
while also containing the reciprocity symmetry expected upon quantization. 
As has been shown here, Hermitian Gravity does successfully retain 
Newton's theory and it is quite 
possible that a more involved complexification scheme, 
rather than an extension to complex Hermitian spaces, 
can be employed to develop a complex gravitational theory
which also correctly reproduces (linearized) general relativity.

%
%

\section*{Appendix A: Constructing Gauge Invariant Potentials}

One constructs the gauge invariant potentials by combining the gauge transformed fields (\ref{gaugetr}) such that the fields $\xi^{\mu}$ cancel out. There are in fact several gauge invariant combinations one can compose
\begin{eqnarray}\label{gicomb}
J_i&=&\partial_i \left(\partial_0 B + 2 \tilde{\psi} - \nabla^2 E\right),\nonumber\\
\bar{J}_{\bar{i}}&=&\partial_{\bar{i}} \left(\partial_{\bar{0}} \bar{B} + 2 \tilde{\psi} - \nabla^2 E\right),\nonumber\\
K&=&2\partial_0 \tilde{\phi} + \frac{1}{2}\nabla^2 \bar{B},\nonumber\\
\bar{K}&=&\partial_{\bar{0}} \tilde{\phi} + \frac{1}{2}\nabla^2 B,\nonumber\\
L&=&2\tilde{\phi}+\nabla^2 E,\nonumber\\
M&=&\nabla^2 \left(\partial_0 E - \frac{1}{2} \bar{B}\right),\nonumber\\
\bar{M}&=&\nabla^2 \left(\partial_{\bar{0}} E - \frac{1}{2} B\right),\nonumber\\
N_i&=&2 \partial_i\left(\tilde{\phi}+\tilde{\psi}+\frac{1}{2}\partial_0 B\right),\nonumber\\
\bar{N}_{\bar{i}}&=&2 \partial_{\bar{i}}\left(\tilde{\phi}+\tilde{\psi}+\frac{1}{2}\partial_{\bar{0}} \bar{B}\right).
\end{eqnarray}
Shortly it will be shown that not all of these quantities are independent of each other. Furthermore, it is not possible to rewrite the vector combinations as derivatives of scalars, i.e. $J_i\neq \partial_i J$, since the quantity $J$ is not gauge invariant. However, the vector combinations do satisfy the symmetry $\partial_j J_i = \partial_i J_j$. Thus the above set of equations represents a (over)complete set of quantities, which we will use to contruct a gauge invariant formulation. The equations of motion (\ref{boxfields}) can be used in conjunction with the gauge invariant combinations (\ref{gicomb}) in order to construct equations that are manifestly gauge invariant. For example, the first and fourth equation~\footnote{In the fourth equation, one makes use of the equation of motion for $E$ only, which corresponds to the terms on the right hand side which do not have a $\delta_{i\bar{j}}$ term.} of (\ref{boxfields}) can be used to produce
\begin{eqnarray}\label{boxLeq}
\frac{1}{6}\Box L &=& \Box \left( \frac{1}{3}\tilde{\phi} + \frac{1}{6}\nabla^2 E\right) = \Box  \psi =\frac{\kappa \rho}{3}.
\end{eqnarray}
This is the first gauge invariant potential. Since in terms of the original fields it is closely related to $\psi$, it shall be called $\Psi=\tilde{\phi} + \frac{1}{2}\nabla^2 E$. Notice that, while the gauge invariant potential is similar to the case of general relativity (\ref{Bardeen:pot}), the equation of motion is quite different from general relativity where it is given by $\Box \Psi_{GR} = \kappa \rho$. The equations of motion for the other gauge invariant combinations are given by~\footnote{The equations for the barred fields follow from complex conjugation.}:
\begin{eqnarray}
\Box J_i &=& \Box \left( \partial_0 \partial_i B + 2\partial_i \tilde{\psi} - \partial_i \nabla^2 E\right)\nonumber\\
&=&-\partial_0 \partial_0 \partial_{\bar{0}} \partial_i B - 2 \partial_0 \partial_{\bar{0}}\partial_i \tilde{\psi} + \partial_0 \partial_{\bar{0}}\partial_i \nabla^2 E -2 \partial_0 \partial_{\bar{0}}\partial_i \tilde{\phi} - \frac{1}{2}\partial_{\bar{0}}\partial_i \nabla^2 \bar{B}\nonumber\\
&=&-\partial_{\bar{0}} \left(\partial_0 J_i + \partial_i K\right),\nonumber\\
\Box N_i &=&\Box \left(2 \partial_i \tilde{\phi} + 2 \partial_i \tilde{\psi} + \partial_0 \partial_i B\right)\nonumber\\
&=&-\frac{1}{2}\partial_{\bar{0}}\partial_i\nabla^2 \bar{B} - 2 \partial_0 \partial_{\bar{0}}\partial_i \tilde{\phi} - \partial_0 \partial_0 \partial_{\bar{0}}\partial_i B - 2 \partial_0 \partial_{\bar{0}} \partial_i \tilde{\psi} + \partial_0 \partial_{\bar{0}}\partial_i \nabla^2 E-2\kappa \partial_i  \rho \nonumber\\
&=&-\partial_{\bar{0}}\left(\partial_0 J_i + \partial_i K\right)-2\kappa \partial_i \rho,\nonumber\\
\Box K &=& \Box \left(2\partial_0 \tilde{\phi} + \frac{1}{2} \nabla^2 \bar{B}\right),\nonumber\\
&=&\partial_0 \nabla^2 \tilde{\psi} + \partial_0 \nabla^2 \tilde{\phi} - \frac{1}{2} \partial_0 \nabla^4 E + \frac{1}{2} \partial_0 \partial_0 \nabla^2 B + \frac{1}{4} \nabla^4 \bar{B}-2\kappa\partial_0  \rho\nonumber\\
&=&\partial_0 \partial_{\bar{i}}J_i + \frac{1}{2} \nabla^2 K-2\kappa\partial_0  \rho,\nonumber\\
\Box M &=&\Box \left(\partial_0 \nabla^2 E - \frac{1}{2} \nabla^2 \bar{B}\right)\nonumber\\
&=&-\partial_0 \nabla^2 \left( \tilde{\psi} + \tilde{\phi} - \frac{1}{2} \nabla^2 E\right) - \frac{1}{2}\partial_0 \partial_0 \nabla^2 B - \frac{1}{4} \nabla^4 \bar{B}\nonumber\\
&=&-\left(\frac{1}{2}\nabla^2 K + \partial_0 \partial_{\bar{i}} J_i \right)\nonumber.
\end{eqnarray}
In the case where $\rho=0$, the two vector and two scalar combinations are, in fact, equivalent under the operation of the d' Alembertian. In other words, $\Box J_i = \Box N_i\rightarrow J_i = N_i + f(z,\bar{z})$, which implies that they are equivalent up to some function $f$ which satisfies $\Box f(z,\bar{z}) =0$, while $\Box \left(K+M\right) = 0$ has a similar solution. Both of these equivalences stem from the fact $\Box L =0$, which allows one to transform $\tilde{\phi}$ into $E$ and vice versa through the the d' Alembertian as can be seen from (\ref{boxLeq}). In terms of dynamics, only three of the gauge invariant combinations (excluding conjugate fields) are then actually independent. $L$ has already been exhausted to form one gauge invariant potential, and thus one must look to $J_i$ and $K$ to construct another. Indeed, this is possible by considering 
\begin{eqnarray}
\Box \left(\partial_{\bar{0}} K + \partial_{\bar{i}} J_i \right) = -2 \kappa \partial_0 \partial_{\bar{0}} \rho.\nonumber
\end{eqnarray}
In terms of the trace-reversed fields this is
\begin{equation}\label{gipsi0}
\Box \left( 2 \partial_0 \partial_{\bar{0}} \tilde{\phi} + \frac{1}{2}\nabla^2 \left(\partial_{\bar{0}}\bar{B}+\partial_0 B \right) + \nabla^2 \tilde{\psi} - \frac{1}{2}\nabla^4 E\right) = 2\kappa \partial_0 \partial_{\bar{0}} \rho.
\end{equation}
By using the relation $\Box L-2\kappa \rho=0$, the $\tilde{\phi}$ term can be re-expressed in terms of $E$ and all that remains is
\begin{equation}\label{gipsi1}
\Box \nabla^2 \left(-2 \partial_0 \partial_{\bar{0}} E + \partial_0 B + \partial_{\bar{0}} \bar{B} + 2 \tilde{\psi} - \nabla^2 E\right) = 0.
\end{equation}
At this point it is useful to simplify the equations of motion for the $B$ field~\footnote{This was not done previously, since it is easier to form gauge invariant combinations by leaving it in the form given in (\ref{eomrev1}).} in (\ref{eomrev1}) by identifying the $B$ containing terms on the right-hand side of the equation as $-\frac{1}{2}\Box \partial_i B$ and thus
\begin{eqnarray}
\Box \partial_i B = 4 \partial_{\bar{0}} \partial_i \left( \tilde{\phi} - \tilde{\psi} + \frac{1}{2} \nabla^2 E\right).\nonumber
\end{eqnarray}
From this, the following relation is easily derived
\begin{eqnarray}
\Box \nabla^2 B = 4 \partial_{\bar{0}} \nabla^2 \left(\tilde{\phi}-\tilde{\psi} + \frac{1}{2}\nabla^2 E\right),\nonumber
\end{eqnarray}
from which one can conclude
\begin{eqnarray}
\Box \nabla^2 \partial_{\bar{0}} \bar{B} = \Box \nabla^2 \partial_0 B,\nonumber
\end{eqnarray}
where it becomes clear that $\partial_0 B$ is a real quantity. Hence this can be introduced into (\ref{gipsi1}) 
\begin{eqnarray}
\Box \nabla^2 \left(-2 \partial_0 \partial_{\bar{0}} E + 2\partial_0 B +  2 \tilde{\psi} - \nabla^2 E\right) = 0.\nonumber
\end{eqnarray}
One last reduction can be made by reconsidering the equation (\ref{gipsi0}). The $2 \partial_0 \partial_{\bar{0}} \tilde{\phi}$ term can be expressed as $-\Box \tilde{\phi} + \nabla^2 \tilde{\phi}$. By plugging in the equation for $\Box \tilde{\phi}$ from (\ref{boxfields}) and using $\Box \nabla^2 \tilde{\phi} = -\frac{1}{2}\Box \nabla^4 E+\kappa \nabla^2 \rho$, one finds
\begin{equation}\label{EBeq}
\Box \nabla^2\left(\partial_0 B - \nabla^2 E\right)=0.
\end{equation}
The above relation permits the omission of two terms in (\ref{gipsi1}) and the final solution reads
\begin{eqnarray}
\Box \nabla^2 \left(-2 \partial_0 \partial_{\bar{0}} E + \partial_0 B +   2\tilde{\psi} \right) = 0.\nonumber
\end{eqnarray}
This equation has the solution~\footnote{For localized distributions, $\nabla^2 g(\vec{z},\vec{\bar{z}})=0$.}
\begin{equation}\label{gipsiim2}
\Box  \left(-2 \partial_0 \partial_{\bar{0}} E + \partial_0 B +  2 \tilde{\psi} \right)  + g(z^0,z^{\bar{0}})= 0.
\end{equation}
However, since the topic of interest concerns localized mass sources, there should be no arbitrary time-dependent functions, because the fields should vanish infinitely far from the source. As a result, $g$ must be a constant which can be fixed to zero. Consequently, the following relation remains
\begin{eqnarray}
\Box  \left(-2 \partial_0 \partial_{\bar{0}} E + \partial_0 B +  2 \tilde{\psi} \right)= 0.\nonumber
\end{eqnarray}
Rewriting this in terms of the original fields, we observe that it contains both contributions from $\phi$ and $\psi$
\begin{eqnarray}
\Box\left(-2 \partial_0 \partial_{\bar{0}} E + \partial_0 B - 4\psi + 2\phi +\nabla^2 E\right)=0.\nonumber
\end{eqnarray}
Using (\ref{boxLeq}), the $\psi$ contribution can be rewritten in terms of $\kappa \rho$:
\begin{equation}\label{gipsiim1}
\Box\left(-2 \partial_0 \partial_{\bar{0}} E + \partial_0 B + 2\phi +\nabla^2 E\right)=\frac{4\kappa\rho}{3},
\end{equation}
which upon inclusion of the relation (\ref{EBeq}) reduces to~\footnote{Strictly speaking, this requires another $\nabla^2$ in (\ref{gipsiim1}), however this is just the $\nabla^2$ that has already been integrated over in (\ref{gipsiim2}).}
\begin{eqnarray}
\Box\left(\phi + \partial_0 B - \partial_0 \partial_{\bar{0}} E\right) = \frac{2\kappa \rho}{3}.\nonumber
\end{eqnarray}
Associating with this expression a gauge invariant quantity $\Phi$ satisfying $\Box \Phi=\frac{2\kappa \rho}{3}$ we have
\begin{eqnarray}
\Phi = \phi + \partial_0\left(B - \partial_{\bar{0}} E\right).\nonumber
\end{eqnarray}
As with the other gauge invariant potential, this one is also strikingly similar to the corresponding $\Phi_{GR}$ in (\ref{Bardeen:pot}). However, the gauge invariant potentials in Hermitian Gravity are quite different than in GR. In general relativity it is found that $\Phi_{GR}=\Psi_{GR}$, but in HG the relation differs by a factor of two, $\Phi=2\Psi$. While in GR one has to solve the equations
\begin{eqnarray}
\Box \Phi_{GR} &=& \Box \left(\phi + \partial_0 \left(B - \partial_0 E\right)\right)=\kappa \rho,\nonumber\\
\Box \Psi_{GR} &=& \Box \psi = \kappa \rho,\nonumber
\end{eqnarray}
in Hermitian Gravity the dynamics are generated by
\begin{eqnarray}
\Box \Phi &=& \Box \left(\phi + \partial_0 \left(B - \partial_{\bar{0}} E\right)\right)=\frac{2\kappa \rho}{3},\nonumber\\
\Box \Psi &=& \Box \psi = \frac{\kappa \rho}{3}.\nonumber
\end{eqnarray}
These are the equations that shall be used to recover the Newtonian limit.

\section*{Appendix B: Constrained Bardeen Potentials}

The holomorphic and anti-holomorphic sectors of the Einstein tensor are given by
\begin{eqnarray}
G_{\mu\nu}&=&\frac{1}{2}\left(\partial_{\mu}\partial^{\bar{\lambda}}\tilde{H}_{\bar{\lambda}\nu}+\partial_{\nu}\partial^{\bar{\lambda}}\tilde{H}_{\bar{\lambda}\mu}\right),\nonumber\\
G_{\bar{\mu}\bar{\nu}}&=&\frac{1}{2}\left(\partial_{\bar{\mu}}\partial^{\lambda}\tilde{H}_{\lambda\bar{\nu}}+\partial_{\bar{\nu}}\partial^{\lambda}\tilde{H}_{\lambda\bar{\mu}}\right).\nonumber
\end{eqnarray}
If the condition $\mbox{\boldmath$\partial^m H_{mn}=0$}$ is satisfied, then $G_{\mu \nu}=G_{\bar{\mu}\bar{\nu}}=0$. To satisfy this condition is not so straightforward however, since in Hermitian Gravity this quantity does not transform under gauge transformations
\begin{equation}\label{dedond}
\mbox{\boldmath$\partial^m H_{mn} \rightarrow \partial^m H_{mn}$}.
\end{equation}
In general relativity, one is able to impose the de Donder gauge $\partial^{\mu}h_{\mu\nu}=0$ using gauge degrees of freedom. In Hermitian Gravity, this is not possible and setting (\ref{dedond}) to zero corresponds to a \emph{physical} choice rather than a gauge choice. Assuming this provides the correct physical picture, this significantly simplifies the Bardeen potential calculations. The Einstein tensor (\ref{EinTensor}) reduces to 
\begin{eqnarray}
G_{\mu\nu}&=&0,\nonumber\\
G_{\bar{\mu}\bar{\nu}}&=&0,\\
G_{\bar{\mu}\nu}&=&-\frac{1}{2}\Box \tilde{H}_{\bar{\mu}\nu},\nonumber\\
G_{\mu\bar{\nu}}&=&-\frac{1}{2}\Box \tilde{H}_{\mu\bar{\nu}}.\nonumber
\end{eqnarray}
From the mixed sectors (holomorphic + anti-holomorphic components), one obtains the following equations of motion from (\ref{Hermitean-Einstein}) for $T_{\bar{\mu}\nu}={\rm diag}\left(-\rho,0,0,0\right)$, a perfect fluid in the rest frame
\begin{eqnarray}
\Box \tilde{\phi} &=&\kappa \rho,\nonumber\\
\Box \partial_i \bar{B}&=&0,\nonumber\\
\Box \left(\tilde{\psi} \delta_{i\bar{j}}-\partial_i\partial_{\bar{j}} E\right)&=&0.\nonumber
\end{eqnarray}
For the holomorphic and antiholomorphic parts, the following equations are generated~\footnote{Here the equations of motion have been written in terms of the gauge invariant combinations for convenience.}
\begin{eqnarray}
G_{00}&=&-\partial_0 K,\nonumber\\
G_{0i}&=&\frac{1}{2}\left(\partial_i K+ \partial_0 J_i\right)=0\nonumber,\\
G_{ij}&=&\partial_j J_i=0,\nonumber
\end{eqnarray}
which impose $K=0$ and $J_i=0$ and similarly for their complex conjugates. From the equation for $K=0$, one can derive
\begin{eqnarray}
K=2\partial_0 \tilde{\phi}+\frac{1}{2}\nabla^2 \bar{B}=6\partial_0 \Psi - M = 0\Rightarrow 6\partial_0 \Psi = M,\nonumber
\end{eqnarray}
while from the equation for $\bar{J}_{\bar{i}}=0$ the following can be gleaned
\begin{eqnarray}
\partial_i \bar{J}_{\bar{i}} = \partial_{\bar{0}}\partial_{\bar{i}} B + 2\partial_{\bar{i}} \tilde{\psi}-\partial_{\bar{i}} \nabla^2 E=\nabla^2 \left(\Phi-2\Psi\right)+\partial_{\bar{0}}M=0\Rightarrow\nabla^2 \left(\Phi-2\Psi\right) = -\partial_{\bar{0}}M.\nonumber
\end{eqnarray}
Combining the two equations and using $\Box \Psi=\frac{\kappa\rho}{3}$, it follows that
\begin{eqnarray}
\nabla^2\left(\Phi + \Psi\right)=\kappa\rho.\nonumber
\end{eqnarray}
 The above equation is in fact completely compatible with the result found in the unconstrained theory
\begin{eqnarray}
\Box\left(\Phi + \Psi\right) = \kappa\rho,\nonumber
\end{eqnarray}
and represents a simplification of it. By constraining $G_{\mu\nu}$ to be zero, the time dependence of the sum of the fields in the equation for the source drops out. This means this particular combination of fields has become static. In other words, if the two gauge invariant potentials are redefined in terms of the sum and difference of the two fields, which are clearly also gauge invariant, then one of the fields becomes static. Hence, under this contraint, the theory is reduced from two dynamical degrees of freedom to just one. Notice, for our consideration of a static, pointlike mass source it does not matter which choice we make!
\pagebreak

\end{document}